\documentclass[aps,prl,twocolumn,showpacs,amsmath,amssymb,superscriptaddress]{revtex4}
\usepackage{graphicx}
\usepackage{amssymb}
\usepackage{natbib}

\begin{document}

\title{Uniaxial pressure effect on the magnetic ordered moment and transition temperatures in BaFe$_{2-x}T_x$As$_2$ ($T=$Co, Ni)}

\author{David W. Tam}
\thanks{These authors contributed equally to this work}
\affiliation{Department of Physics and Astronomy, Rice University, Houston, Texas 77005, USA}
\author{Yu Song}
\thanks{These authors contributed equally to this work}
\affiliation{Department of Physics and Astronomy, Rice University, Houston, Texas 77005, USA}
\author{Haoran Man}
\affiliation{Department of Physics and Astronomy, Rice University, Houston, Texas 77005, USA}
\author{Sky C. Cheung}
\affiliation{Department of Physics, Columbia University, New York, New York 10027, USA}
\author{Zhiping Yin}
\affiliation{Center for Advanced Quantum Studies and Department of Physics, Beijing Normal University, Beijing 100875, China}
\author{Xingye Lu}
\affiliation{Department of Physics and Astronomy, Rice University, Houston, Texas 77005, USA}
\author{Weiyi Wang}
\affiliation{Department of Physics and Astronomy, Rice University, Houston, Texas 77005, USA}
\author{Benjamin A. Frandsen}
\affiliation{Department of Physics, Columbia University, New York, New York 10027, USA}
\author{Lian Liu}
\affiliation{Department of Physics, Columbia University, New York, New York 10027, USA}
\author{Zizhou Gong}
\affiliation{Department of Physics, Columbia University, New York, New York 10027, USA}
\author{Takashi U. Ito}
\affiliation{Advanced Science Research Center, Japan Atomic Energy Agency, Tokai, Ibaraki 319-1195, Japan}
\author{Yipeng Cai}
\affiliation{Department of Physics and Astronomy, McMaster University, Hamilton, Ontario L8S 4M1, Canada}
\author{Murray N. Wilson}
\affiliation{Department of Physics and Astronomy, McMaster University, Hamilton, Ontario L8S 4M1, Canada}
\author{Shengli Guo}
\affiliation{Department of Physics, Zhejiang University, Hangzhou 310027, China}
\author{Keisuke Koshiishi}
\affiliation{Department of Physics, Univeristy of Tokyo, 7-3-1 Hongo, Bunkyo-Ku, Tokyo 113, Japan}
\author{Wei Tian}
\affiliation{Quantum Condensed Matter Division, Oak Ridge National Laboratory, Oak Ridge, Tennessee 37831, USA}
\author{Bassam Hitti}
\affiliation{TRIUMF, Vancouver, British Columbia, V6T2A3, Canada}
\author{Alexandre Ivanov}
\affiliation{Institut Laue-Langevin, 71 avenue des Martyrs, 38000 Grenoble, France}
\author{Yang Zhao}
\affiliation{NIST Center for Neutron Research, National Institute of Standards and Technology, Gaithersburg, Maryland 20899, USA}
\affiliation{Department of Materials Science and Engineering, University of Maryland, College Park, Maryland 20742, USA}
\author{Jeffrey W. Lynn}
\affiliation{NIST Center for Neutron Research, National Institute of Standards and Technology, Gaithersburg, Maryland 20899, USA}
\author{Graeme M. Luke}
\affiliation{Department of Physics and Astronomy, McMaster University, Hamilton, Ontario L8S 4M1, Canada}
\author{Tom Berlijn}
\affiliation{Center for Nanophase Materials Sciences and Computer Science and Mathematics Division, Oak Ridge National Laboratory, Oak Ridge, Tennessee 37831-6494, USA}
\author{Thomas A. Maier}
\affiliation{Center for Nanophase Materials Sciences and Computer Science and Mathematics Division, Oak Ridge National Laboratory, Oak Ridge, Tennessee 37831-6494, USA}
\author{Yasutomo J. Uemura}
\affiliation{Department of Physics, Columbia University, New York, New York 10027, USA}
\author{Pengcheng Dai}
\email{pdai@rice.edu}
\affiliation{Department of Physics and Astronomy, Rice University, Houston, Texas 77005, USA}
\affiliation{Center for Advanced Quantum Studies and Department of Physics, Beijing Normal University, Beijing 100875, China}

\date{\today}
\pacs{74.70.Xa, 75.30.Gw, 78.70.Nx}

\begin{abstract}
We use neutron diffraction and muon spin relaxation to study the effect of in-plane uniaxial pressure on the antiferromagnetic (AF) orthorhombic phase in BaFe$_2$As$_2$ and its Co- and Ni-substituted members near optimal superconductivity. In the low temperature AF ordered state, uniaxial pressure necessary to detwin the orthorhombic crystals also increases the magnetic ordered moment, reaching an 11\% increase under 40 MPa for BaFe$_{1.9}$Co$_{0.1}$As$_2$, and a 15\% increase for BaFe$_{1.915}$Ni$_{0.085}$As$_2$. We also observe an increase of the AF ordering temperature ($T_N$) of about 0.25 K/MPa in all compounds, consistent with density functional theory calculations that reveal better Fermi surface nesting for itinerant electrons under uniaxial pressure. The doping dependence of the magnetic ordered moment is captured by combining dynamical mean field theory with density functional theory, suggesting that the pressure-induced moment increase near optimal superconductivity is closely related to quantum fluctuations and the nearby electronic nematic phase.
\end{abstract}

\maketitle

Understanding the behavior of magnetism in iron superconductors continues to be an important topic in modern condensed matter physics because spin excitations may mediate electron pairing for high-temperature superconductivity \cite{kamihara,cruz,qhunag,WYRF09,mgkim,scalapino,dai}. Some of the earliest work in this field determined that iron pnictides such as LaFeAsO \cite{kamihara,cruz} and BaFe$_2$As$_2$ [Figs. 1(a) and 1(b)] \cite{qhunag,WYRF09,mgkim} form static stripe antiferromagnetic (AF) order at $T_N$ preceded by a tetragonal-to-orthorhombic structural transition of the lattice at $T_s$ ($T_N\leq T_s<300$ K). While AF order may be a spin-density wave (SDW) from nesting of hole and electron Fermi surfaces at the $\Gamma$ and $X$ points in the one-iron Brillouin zone [Figs. 1(c) and 1(d)], respectively \cite{IIMazin,hirschfeld,chubukov,supplementary}, it may also originate from localized moments on individual Fe sites \cite{si,cfang,xuc,Basov11}.

Upon hole doping to form Ba$_{1-x}$Na$_{x}$Fe$_{2}$As$_{2}$, a magnetically ordered state with restored tetragonal symmetry is found near optimal superconductivity, replacing the stripe AF ordered state \cite{Avci13,Waber}. M$\rm \ddot{o}$ssbauer spectroscopy experiments find real space modulation of magnetic moments on Fe sites, thus conclusively establishing that the magnetic order with tetragonal symmetry is a SDW from itinerant electrons \cite{Allred16}. For electron doped BaFe$_{2-x}T_x$As$_2$ ($T=$Co, Ni), long-range commensurate stripe magnetic order in BaFe$_{2}$As$_{2}$ evolves into short-range incommensurate magnetic order near optimal superconductivity \cite{Pratt12,HQluo12}, due possibly to a SDW order \cite{Pratt12} or a cluster spin glass \cite{Dioguardi13,Dioguardi,XYLu14a}. While these results suggest a rich variety of magnetic ground states for superconducting iron pnicitdes, nematic order has recently been identified as a unifying feature near optimal superconductivity, seen via the resistivity anisotropy induced by in-plane mechanical strain \cite{KCPK16}. Since nematic order couples linearly to anisotropic strain of the same symmetry \cite{FeSc12,anna}, a determination of the effect of uniaxial pressure on magnetism of iron pnictides 
should reveal if the observed nematic susceptibility \cite{KCPK16} is associated with magnetic order and spin excitations.

\begin{figure*}[t]
\includegraphics[scale=.33]{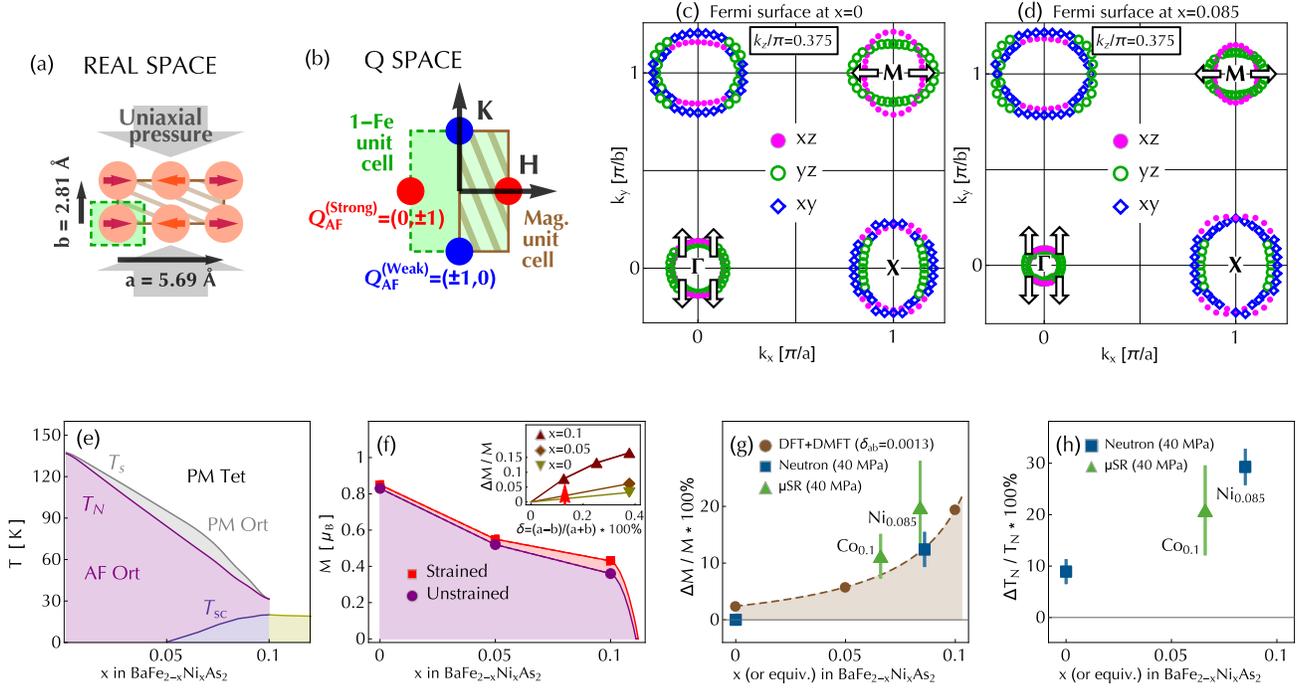}
\caption{(a-b) Real-space and ${\bf Q}$-space configuration in the AF state, showing the majority domain Bragg reflections $\mathbf{Q}_\text{strong}=(\pm 1,0)$ in red dots for pressure applied along the $b$-axis, and the minority domain Bragg reflections $\mathbf{Q}_\text{weak}=(0,\pm 1)$ as blue dots. 
(c-d) Schematic illustration of the uniaxial-pressure-induced Fermi surface distortion parallel to the $ab$-plane (at $k_z$=$0.375\pi/c$) for BaFe$_2$As$_2$ (c) and BaFe$_{1.915}$Ni$_{0.085}$As$_2$ (d). Arrows indicate the direction of distortion in this $k_z$ plane for the uniaxially-strained case, which is much smaller than the thickness of the markers. Coloring shows the dominant orbital character as indicated in the inset.
(e) Experimental phase diagram of BaFe$_{2-x}$Ni$_x$As$_2$ \cite{LGZL13}. 
(f) Electron-doping dependence of the ordered magnetic moment with ($M$; red) and without ($M_0$; purple) uniaxial pressure obtained from a combined DFT/DMFT calculation. The inset shows $\delta=(a-b)/(a+b)$ dependence of $M/M_0$ at $x=0,0.05$, and 0.1. 
(g) Theoretical/experimental results demonstrating the enhancement of $\Delta M/M$ as $M$ decreases on 
approaching optimal doping \cite{supplementary}.
(h) Enhancement of $\Delta T_N/T_N$, as $T_N$ decreases on approaching optimal doping.
}
\end{figure*}

Without uniaxial strain, BaFe$_{2-x}T_x$As$_2$ forms twinned domains in the orthorhombic state, with AF Bragg reflections occurring at $(\pm 1,0,L)$ and $(0,\pm 1,L)$ ($L=1,3,5,\ldots$) \cite{dai}. Uniaxial pressure has been used to mechanically detwin single crystals by compressing along one axis of the orthorhombic lattice, creating a preferred orientation for microscopic domains \cite{jhchu,matanatar,fisher}. However, even the modest amount of pressure necessary for detwinning ($\sim$10 MPa) also induces a significant ($\sim$1-2 K) upward shift of $T_N$ in electron-doped BaFe$_2$As$_2$ \cite{dhital,Dhital14,YSong13,xylu14,man}, and changes $T_s$ into a crossover with orthorhombic lattice distortion at all temperatures \cite{XYLu15}. One study combining a phenomenological Ginsburg-Landau model with density functional theory (DFT) calculations under uniaxial strain suggest that the magnetic ordered moment of BaFe$_2$As$_2$ decreases under pressure \cite{Tomic}, in agreement with the experimentally observed decrease in the combined magnetic scattering at $(\pm 1,0,L)$ and $(0,\pm 1,L)$ \cite{dhital,Dhital14}. However, the bulk-averaged nature of neutron measurements cannot distinguish a change in magnetic volume fraction from a changing ordered moment. To conclusively determine the effect of uniaxial strain on magnetic order, one must combine neutron scattering with a probe such as muon spin relaxation ($\mu$SR) \cite{GABB09,ABBC08}.

\begin{figure}[h]
\includegraphics[scale=.25]{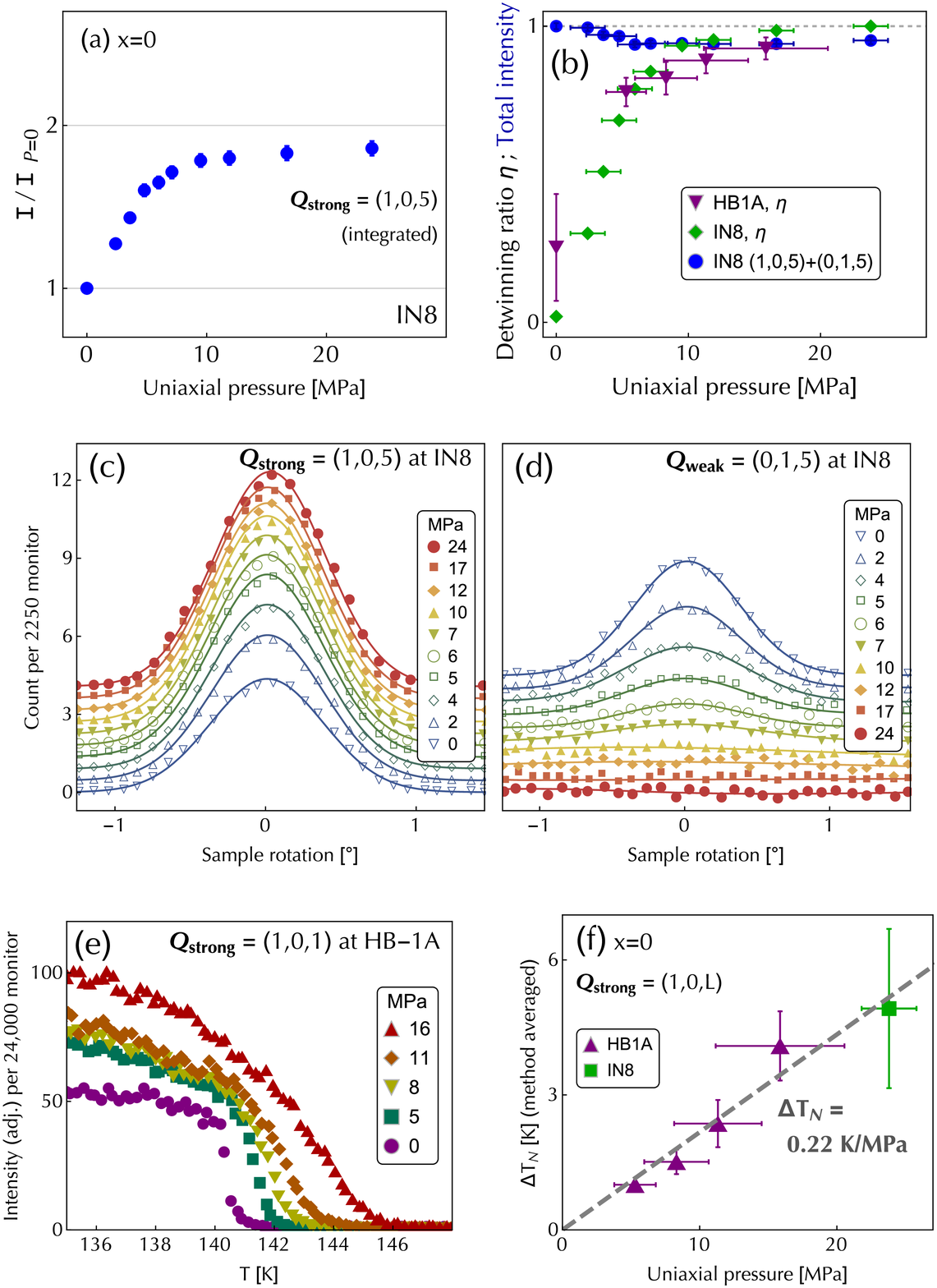}
\caption{ 
(a) Intensity change of the $\mathbf{Q}_\text{strong}=(1,0,5)$ peak 
with uniaxial pressure on BaFe$_2$As$_2$ at IN8.
(b) Detwinning ratio $\eta=(I_\text{strong}-I_\text{weak})/(I_\text{strong}+I_\text{weak})$ for both HB-1A and IN8 experiments, and the total intensity (circles) $I_\text{strong}+I_\text{weak}$ at IN8 which remains conserved, up to small corrections that we attribute to the extinction effect. (c-d) Rocking curves at $(1,0,5)$ and $(0,1,5)$ measured at IN8 at $T=90$ K, demonstrating nearly 100\% detwinning above 10 MPa. The solid lines are fits to a single Gaussian peak.
(e) Temperature dependence of $\mathbf{Q}_\text{strong}=(1,0,1)$ on warming. For clarity, the vertical scale of each scan has been slightly adjusted to represent the total magnetic scattering intensity, using the integrated intensity of a rocking scan measured immediately prior to warming.
(f) Shift in AF ordering temperature $\Delta T_N = T_N^\text{(P)}-T_N^\text{(P=0)}$ ($T_N^\text{(P=0)} \approx 140$ K) for BaFe$_2$As$_2$ \cite{supplementary}.
}
\end{figure}

In this paper, we use neutron diffraction, $\mu$SR, and combined DFT and dynamical mean field theory (DMFT) calculations [Figs. 1(f), 1(g), and 1(h)] \cite{ZPYin11,ZPYin14} to study the effect of uniaxial pressure on the AF phase transition in BaFe$_{2-x}T_x$As$_2$. In BaFe$_2$As$_2$, the sample achieves nearly 100\% detwinning for pressures above $\sim$10 MPa [Fig. 2(b)], while the ordered magnetic moment remains constant or reduces very slightly, resulting in a doubling of the neutron magnetic scattering intensity [Fig. 2(a)] from the majority-domain Bragg reflections $\mathbf{Q}_\text{strong}=(\pm 1,0)$ and elimination of magnetic scattering from the minority domains $\mathbf{Q}_\text{weak}=(0,\pm 1)$ [Figs. 2(c) and 2(d)]. By contrast, in BaFe$_{1.915}$Ni$_{0.085}$As$_2$, the scattering intensity at $\mathbf{Q}_\text{strong}=(\pm 1,0)$ more than doubles for pressures greater than $\sim$30 MPa [Fig. 3(a)]. We also use $\mu$SR to show the magnetic volume fraction is not changing while the internal magnetic field at the muon site is increasing, conclusively establishing that the magnetic ordered moment is increasing under uniaxial strain.
We also find the magnetic ordering temperature for different iron pnictides \cite{HQluo12} increases at approximately the same rate $T_N^\text{(P)}-T_N^\text{(P=0)} \approx 0.24$ K/MPa [Fig. 2(f) and 3(b)], consistent with our DFT calculations of the 
nesting condition under pressure \cite{supplementary}. Our $\mu$SR measurements also demonstrate that the AF phase transition is broadened, indicating that the internal uniaxial strain has a distribution for a nominally constant induced stress.

\begin{figure}[h!]
\includegraphics[scale=.25]{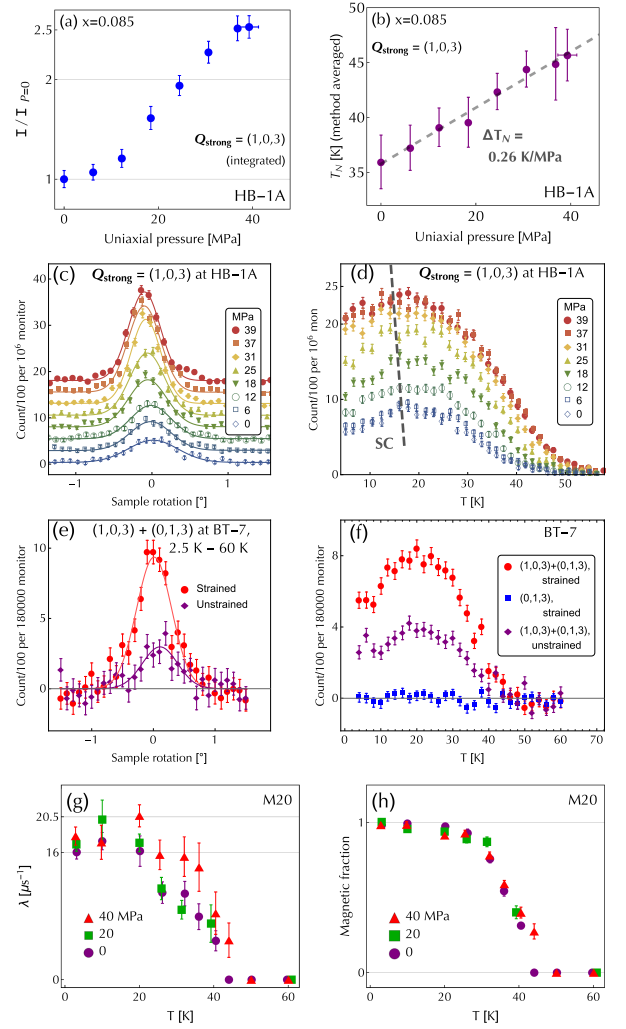}
\caption{
(a) Intensity change of $\mathbf{Q}_\text{strong}=(1,0,3)$
with uniaxial pressure  on BaFe$_{1.915}$Ni$_{0.085}$As$_2$ at HB-1A. 
(b) Shift in AF ordering temperature $\Delta T_N = T_N^\text{(P)}-T_N^\text{(P=0)}$ ($T_N^\text{(P=0)} \approx$35 K).
(c) Rocking curves at $\mathbf{Q}_\text{strong}=(1,0,3)$, $T = 20$ K, measured at HB-1A in the $[H,0,L]$ scattering plane.
(d) Temperature scans at HB-1A. The dashed line in indicates the superconducting region ($T_c^\text{(40 MPa)}-T_c^\text{(0)} \approx -3 \text{ K}$) \cite{supplementary}.
(e) Temperature dependence of the magnetic scattering intensity at $\mathbf{Q}_\text{strong}=(1,0,3)$ and $\mathbf{Q}_\text{weak}=(0,1,3)$ with and without pressure, measured in-situ at BT-7.
(f) Combined rocking scans at $(1,0,3)$ and $(0,1,3)$. Data at 60 K has been subtracted from each ${\bf Q}$ position before combining the data.
(g) Fast muon relaxation rate $\lambda$ \cite{supplementary}.
(h) Fast relaxing fraction $V_M$, demonstrating constant magnetic sample volume for all pressures below 35 K.
}
\end{figure}

For our combined DFT and DMFT calculations in the collinear AF state, we use the experimental crystal structure and the same Coulomb interactions used previously for BaFe$_2$As$_2$ \cite{ZPYin11,ZPYin14}. We find that uniaxial strain does indeed enhance the ordered moment for BaFe$_{1.915}$Ni$_{0.085}$As$_2$ (Fig. 3), and that the effect becomes larger for samples close to optimal superconductivity. The inset in Figure 1(f) shows the dependence of the ordered magnetic moment on doping ($x$) and distortion ($\delta=(a-b)/(a+b)$), which supports a much larger increase of the magnetic moment in the doped compounds for fixed $\delta=0.38 \%$, corresponding to about 30 MPa in the parent compound. In Figure 1(f), we fix the distortion $\delta=0.0013$ corresponding to about 10 MPa in the parent compound \cite{XYLu15}, and plot the ordered moment with ($M$ at $a>b$) and without uniaxial pressure ($M_0$ at $a=b$). Using these results, in Fig. 1(g) we define a susceptibility $\chi=[(M-M_0)/(M+M_0)]/[(a-b)/(a+b)]$, which is small in the parent compound but shows diverging behavior near $x=0.1$ (dashed line). Because fixing the distortion $\delta=0.0013$ corresponds to a greater applied uniaxial pressure for the doped samples due to the larger bulk modulus in those compounds \cite{anna}, the overplotted experimental points from neutron and $\mu$SR measurements may underestimate the moment increase relative to the calculated points. Nevertheless, we find clear similarity between the theoretical and experimental results, implying that the divergence of the nematic susceptibility \cite{KCPK16} have a magnetic origin.

Figure 1(e) shows the schematic phase diagram of BaFe$_{2-x}$Ni$_x$As$_2$, where Fig. 1(a) and 1(b) show the AF unit cell and Bragg peak positions in reciprocal space, respectively \cite{dai}. By gradually applying uniaxial pressure along the $b$-axis direction of the orthorhombic structure, the sample becomes increasingly detwinned, which enhances the $\mathbf{Q}_\text{strong}=(\pm 1,0)$ Bragg peak while the $\mathbf{Q}_\text{weak}=(0,\pm 1)$ peaks become extinguished [red and blue dots in Fig. 1(b)] \cite{dhital,Dhital14,YSong13,xylu14}. The striped brown boxes in Fig. 1(a-b) represent the low-temperature magnetic unit cell and its equivalent area in $\bf Q$ space, while in green we show the configurations for a single Fe ion \cite{supplementary}. 

Figure 2 summarizes our experimental results on the uniaxial pressure effect in BaFe$_2$As$_2$. In Fig. 2(c) and 2(d), we show how the Bragg peak intensity becomes redistributed from the weak side [$\mathbf{Q}_\text{weak}=(0,\pm1,L)$] to the strong side [$\mathbf{Q}_\text{strong}=(\pm1,0,L)$], resulting in a near doubling of the strong reflection intensity [Fig. 2(a)] while the total intensity remains conserved [Fig. 2(b)]. Consistent with earlier work \cite{dhital,Dhital14}, we find that pressure both enhances and broadens the magnetic phase transition. The rocking curves [Fig. 2(c-d)] at $(1,0,5)$ and $(0,1,5)$ were collected at IN8 at 90 K \cite{man}. The pressure dependence of the detwinning ratio, defined as $\eta=(I_{10}-I_{01})/(I_{10}+I_{01})$, and total scattering intensity $I_\text{total} = (I_{10}+I_{01})/(I_{10}+I_{01})_{P=0}$ from both experiments are also included in Fig. 2(b), which shows a very small reduction in $I_\text{total}$ between 0 and 5 MPa. 
Fig. 2(e-f) shows the effect of pressure on the magnetic ordering temperature \cite{supplementary}.

Figure 3 summarizes the uniaxial pressure dependence in BaFe$_{1.915}$Ni$_{0.085}$As$_2$. In a twinned sample, the elastic magnetic scattering intensity at $(1,0,3)$ should be equal to that at $(0,1,3)$, ignoring absorption and other instrumentation effects \cite{supplementary}. If magnetic moments do not react to the applied uniaxial pressure, one would expect the scattering intensity for a fully detwinned sample to at most double at the $(1,0,3)$ position, such as in Fig. 2(a) for BaFe$_2$As$_2$. Figure 3(a) shows this is not the case: the scattering intensity at 20 K rises by a factor of $\sim$2.5 from 0 to 40 MPa. Considering the detwinning effect, this corresponds to a 25\% increase in the scattering cross-section, which is proportional to the squared magnetic ordered moment ($M^2$). Fig. 3(c-d) shows the temperature dependence of the strong Bragg reflection, measured in the $[H,0,0]\times [0,0,L]$ scattering plane, which decreases below 20 K due to the onset of superconductivity \cite{Pratt,Christianson09}. There is also a slight decrease in $T_c$ under uniaxial pressure \cite{Hardy,supplementary}. Figure 3(e-f) shows that the combined scattering from $\mathbf{Q}_\text{strong}$ and $\mathbf{Q}_\text{weak}$
nearly doubles in intensity from zero to large uniaxial pressure ($>50$ MPa) \cite{supplementary}. Using these data and the measured magnetic moment of $M=0.08\ \mu_B$/Fe for BaFe$_{1.915}$Ni$_{0.085}$As$_2$ \cite{HQluo12}, we estimate an increase in magnetic moment from 0.08 $\mu_B$/Fe to 0.092 $\mu_B$/Fe.

To disentangle the magnetic volume fraction from the ordered moment in BaFe$_{1.915}$Ni$_{0.085}$As$_2$, we carried out $\mu$SR measurements using the same crystal and under the same conditions \cite{supplementary}. We implanted muons with the spin polarization along the sample $a$-axis and observed fast relaxation due to a build-up of the static internal magnetic field. From the ZF-$\mu$SR time spectra in \cite{supplementary}, we obtained the fast relaxation rate $\lambda$ in Fig. 3(g), which is linearly proportional to the size of the local static magnetic moment ($M$), together with the volume fraction $V_M$ of magnetically ordered regions in Fig. 3(h). In addition to confirming the pressure-induced $T_N$ increase [Fig. 1(h)], we find that the relaxation rates at $T = 20$ K and $T = 3$ K exhibit a $\sim$10-20\% increase from 2.5 MPa to 40 MPa \cite{supplementary}, and see a decreasing trend in $\lambda$ below $T_c$. These $\mu$SR results are consistent with the neutron results, where the $\sim$25\% increase from 2.5 MPa to 40 MPa (after detwinning is considered) is seen in the scattering intensity [Fig. 3(a)], proportional to $M^2$. The fraction of the sample exhibiting magnetic order [Fig. 3(h)] is essentially 100\% below $T = 35$ K with no dependence on uniaxial pressure, while it is slightly larger and broader in the region $30 < T < 50$ K under pressure, consistent with 
the results of neutron scattering [Fig. 2(e), 3(d), 3(f), and 3(g)] \cite{dhital,Dhital14}.

\begin{figure}[t]
\includegraphics[scale=.25]{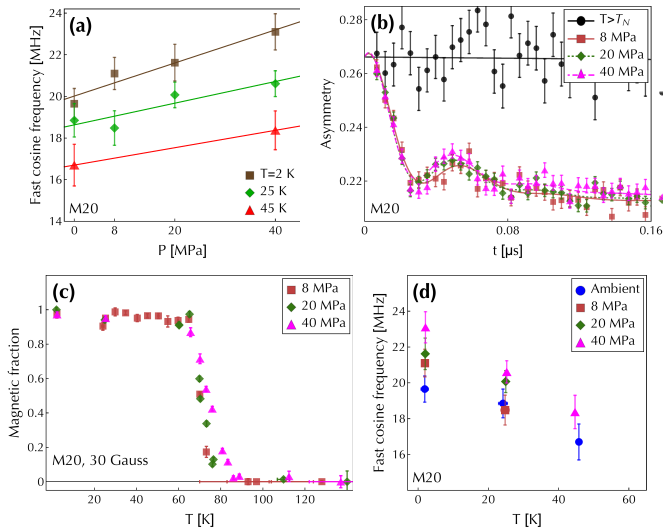}
\caption{
(a) Increase of the oscillation frequency in the fast-relaxing portion of the muon decay asymmetry, for $T=2$ K (superconducting), 25 K, and 45 K for BaFe$_{1.9}$Co$_{0.1}$As$_2$ \cite{supplementary}.
(b) Muon decay asymmetry at 25 K as a function of uniaxial pressure. The fraction of muons remaining polarized in this short time window are those landing in the nonmagnetic sample holder.
(c) Fast relaxing fraction of muons in a 30 G transverse magnetic field, demonstrating a constant magnetic sample volume below $T_N$=65 K, as well as a broadening of the magnetic transition under pressure and an increase in $T_N$.
(d) Data in (a) plotted as an order parameter.
}
\end{figure}

We carried out similar $\mu$SR measurements on BaFe$_{1.9}$Co$_{0.1}$As$_2$ ($T_N=70$ K). In a zero field environment, we observe oscillations in the time spectra that indicate the presence of long-range order, different from the spectra in BaFe$_{1.915}$Ni$_{0.085}$As$_2$ \cite{supplementary} where the larger doping causes the magnetic fields to lose coherence \cite{XYLu15,ABBC08,GABB09,WABB09,WABC14}. Assuming two muon stopping sites in BaFe$_{2-x}$Co$_{x}$As$_2$ \cite{WABC14}, we fit the short-time coherent relaxation with a two-cosine function, separately fitting both frequencies for each temperature and uniaxial pressure, while globally constraining the other free parameters such as relaxation rates \cite{supplementary}. Figure 4(a) shows the uniaxial pressure evolution of the fast cosine frequency in the muon decay time-spectra [Fig. 4(b)] for several temperatures. Although 2 K is below $T_c$, the magnetic field increases at least as much as at 25 K, indicating that the magnetic phase may not be fully saturated at 25 K. To measure the magnetic volume fraction [Fig. 4(c)], we apply a weak (30 G) external magnetic field in order to clearly distinguish the fast relaxing component from the long paramagnetic oscillations. The ordered phase volume saturates below $\sim 65$ K for all pressures, while for higher temperatures the magnetic transition is broadened significantly for BaFe$_{1.9}$Co$_{0.1}$As$_2$ as it was with BaFe$_{1.915}$Ni$_{0.085}$As$_2$ [Fig. 3(h)].

As the doping increases and the ordered magnetic moment correspondingly decreases, it is natural to expect that quantum fluctuations, including those related to a nematic quantum critical point \cite{KCPK16}, become important. Here we have demonstrated that an in-plane symmetry-breaking field is more effective in enhancing the ordered magnetic moment 
for nearly optimally doped iron pnictides. The corresponding decrease in $T_c$ is consistent with the view that by slightly increasing the orthorhombicity, the doped system effectively shifts leftward on the phase diagram (toward the parent compound), and demonstrates that in-plane lattice distortion is a mechanism by which magnetism directly competes with superconductivity. Since the doped compounds react much more sensitively to uniaxial pressure, the enhanced magnetic properties demonstrate the sensitivity of magnetism to quantum fluctuations near optimal superconductivity, suggesting that nematic order is associated with both magnetism and optimal superconductivity in iron pnictides \cite{KCPK16,FeSc12}.

The neutron scattering work at Rice is supported by the U.S. NSF-DMR-1362219 and DMR-1436006 (P.D.). The RPA calculations at Rice/ORNL is supported by NSF-DMR-1308603 (T.A.M. and P.D.). The materials synthesis efforts at Rice is supported by the Robert A. Welch Foundation Grant Nos. C-1839 (P.D.). The research at ORNL was sponsored by the Scientific User Facilities Division, Office of BES, U.S. DOE. The DFT and Wannier function calculations were conducted at the Center for Nanophase Materials Sciences, which is a DOE's Scientific User Facility (T.B.). Work at Columbia and TRIUMF is supported by NSF-DMREF DMR-1436095 and DMR-1610633 (Y.J.U.), JAEA Reimei project and Friends of U. Tokyo Inc.

\clearpage

\section{Supplementary information.}
\setcounter{figure}{0}
\renewcommand{\figurename}{SFigure }

\subsection{Neutron scattering experimental details.}

\begin{figure*}[t]
\includegraphics[scale=.4]{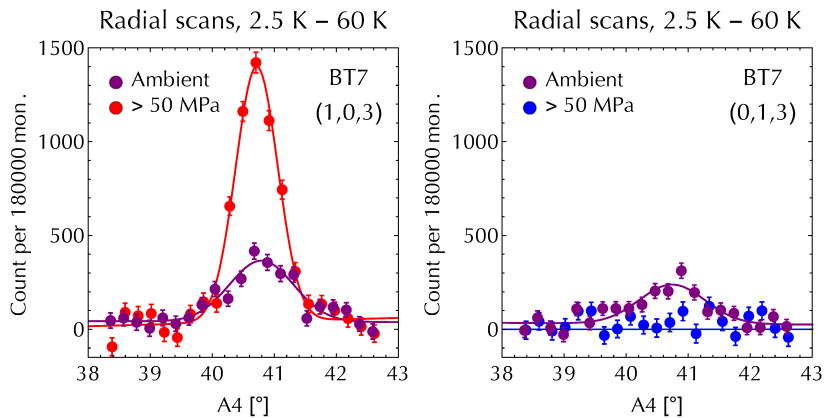}
\caption{Neutron scattering measurements on BaFe$_{1.915}$Ni$_{0.085}$As$_2$ at zero and $>50$ MPa uniaxial pressure. 
Radial scans (L scans) across the (1,0,3) [strong, left panel] and (0,1,3) [weak, right panel] magnetic reflections at BT7, demonstrating the evolution from nearly complete crystal twinning to 100\% detwinning under uniaxial pressure.}
\label{bt7radialplotgrid2}
\end{figure*}

Elastic neutron scattering measurements on BaFe$_2$As$_2$ were carried out at the HB-1A and IN8 spectrometers, respectively at the High Flux Isotope Reactor (HFIR) at Oak Ridge National Laboratory, Oak Ridge, Tennessee, and the Institut Laue-Langevin, Grenoble, France. Measurements on BaFe$_{1.915}$Ni$_{0.085}$As$_2$ were conducted at HB-1A and at the BT-7 triple-axis spectrometer at the NIST Center for Neutron Research, Gaithersburg, Maryland \cite{2bt7}. The $\mu$SR measurements were carried out on the LAMPF spectrometer at the M20D surface muon channel at TRIUMF, Vancouver, Canada. The methods and apparatus used to apply pressure are somewhat different between the experiments, as described below. The wave vector \textbf{Q} in three-dimensional reciprocal space in \AA$^{-1}$ is defined as ${\bf Q}=H {\bf a^\ast}+K{\bf b^\ast}+L{\bf c^\ast}$, where $H$, $K$, and $L$ are Miller indices and ${\bf a^\ast}=\hat{{\bf a}}2\pi/a, {\bf b^\ast}=\hat{{\bf b}} 2\pi/b, {\bf c^\ast}=\hat{{\bf c}}2\pi/c$ are reciprocal lattice units (rlu). In the low-temperature AF orthorhombic phase of BaFe$_2$As$_2$, $a\approx 5.69$ \AA, $b\approx 5.61$ \AA, and $c\approx 12.953$ \AA\ \cite{2qhunag}.

We prepared high quality single crystals of BaFe$_{2-x}T_x$As$_2$ using the self-flux method. The samples were cut into square shapes approximately $10\times10\times 1$ mm in size along the orthorhombic $a/b$ directions using a fine-precision wire saw. Different mechanical devices were used to apply uniaxial pressure, as described below and previously \cite{2xylu14,2man,2XYLu15}.

In all neutron experiments, the integrated intensity was determined from fits to rocking curves using a Gaussian function on a linear background, except as noted below. The error bars for integrated intensity were computed by letting each fit parameter vary over its confidence interval as determined by the fitting algorithm, and determining the global maximum and minimum, which is used as the value of the error bar. Horizontal error bars reflect our best estimate of the applied uniaxial pressure. Temperature scans were collected under warming controlled by a temperature sensor located nearby the sample and a heating element. Nuclear (structural) peaks were measured for all experiments and found not to evolve with pressure to very high precision. Triple-axis spectrometers used in the present experiments do not have the resolution capable of resolving the twinned orthorhombic peaks.

For experiments at HB-1A on BaFe$_2$As$_2$, the crystal was aligned in the $[1,-1,0]\times [1,1,2]$ scattering plane (orthorhombic notation), allowing access to both (1,0,1) and (0,1,1) magnetic Bragg peaks. The sample was mounted in an aluminum can filled with helium buffer gas and connected to a closed-cycle refrigerator (CCR). Uniaxial pressure was applied by compressing a spring at room temperature and then cooling the apparatus with no further changes to the spring. Rocking curves were collected at base temperature as well as above and below $T_N \approx$ 140 K. In this experiment, we fitted the rocking scans to two Gaussian functions in order to accommodate the small minority domain we identified from the nuclear Bragg peak rocking curves. The two Gaussian functions were separated by fixed width for each Bragg position. After reaching base temperature, the sample was warmed to 120 K to collect temperature scans across $T_N$ at one magnetic Bragg peak; then the sample was cooled to roughly 100 K before repeating the scan at the opposite twin position, and again for the (-2,2,0) nuclear peak. The temperature scans across $T_N$ were collected under a relatively fast warming rate of 1 K per 75 seconds.

At IN8, the BaFe$_2$As$_2$ crystal was aligned in the $[-1,0,-5]\times [0,1,5]$ plane using the combined Flat-Cone rotation of the sample and detector as in Ref. \cite{2MLCZ15}. The sample environment was a standard orange (liquid helium) cryostat. Uniaxial pressure was applied by adjusting the spring length from a dial micrometer located outside the cryostat, using a specially designed in-situ control rod \cite{2MLCZ15}. The virgin sample was cooled from 300 to 90 K and pressure was systematically increased to the nominal values shown in Fig. 2 by turning the dial micrometer. The maximum pressure scans were measured first, and the pressure was released in steps back into the fully twinned scenario. We do not expect the order of scans to have any effect on the data. We note that using density functional theory, Tomi{\'c} {\it et al.} \cite{2TJFV13} calculated a critical pressure of nearly 200 MPa required to flip the stripe direction of a microscopic magnetic domain in the low-temperature state; here we observe 100\% detwinning with less than 10 MPa. At IN8, the rocking curves were collected by counting for 4 seconds per point. For both the HB-1A and IN8 experiments, the pressure values were calibrated in absolute units using the compressed distance and the manufacturer's measurement of the spring constant at room temperature. The systematic errors stemming from the temperature dependence of the spring are unknown, but are not expected to be large.

At BT-7, we aligned the BaFe$_{1.915}$Ni$_{0.085}$As$_2$ crystal in the $[1,-1,0]\times [1,1,6]$ scattering plane, allowing access to (1,0,3) and (0,1,3), which have higher magnetic scattering intensity that (1,0,1) and (0,1,1). We did not use a spring, but simply tightened a piece of aluminum against the sample edge with a large amount of force applied by a thumbscrew. The absolute pressure is not known although we believe it is larger than 50 MPa. In SFigure \ref{bt7radialplotgrid2}, we show the evolution of the strong and weak Bragg peaks in BaFe$_{1.915}$Ni$_{0.085}$As$_2$ as observed in the BT7 experiment. We note the very large increase in the strong reflection and the complete loss of signal in the weak reflection. Due to the use of the [1,0,3]$\times$[0,1,3] scattering plane, these radial scans do not have fixed $L$ values.

The HB-1A measurements on BaFe$_{1.915}$Ni$_{0.085}$As$_2$ at HB-1A using the pneumatic uniaxial pressure device described below, mounted vertically (pressure applied downward) in a standard orange (liquid helium) cryostat. In this case, the scattering plane is $[1,0,0]\times [0,0,1]$.

\subsection{Determining $T_N$ from neutron data.}

\begin{figure*}[t]
\includegraphics[scale=.4]{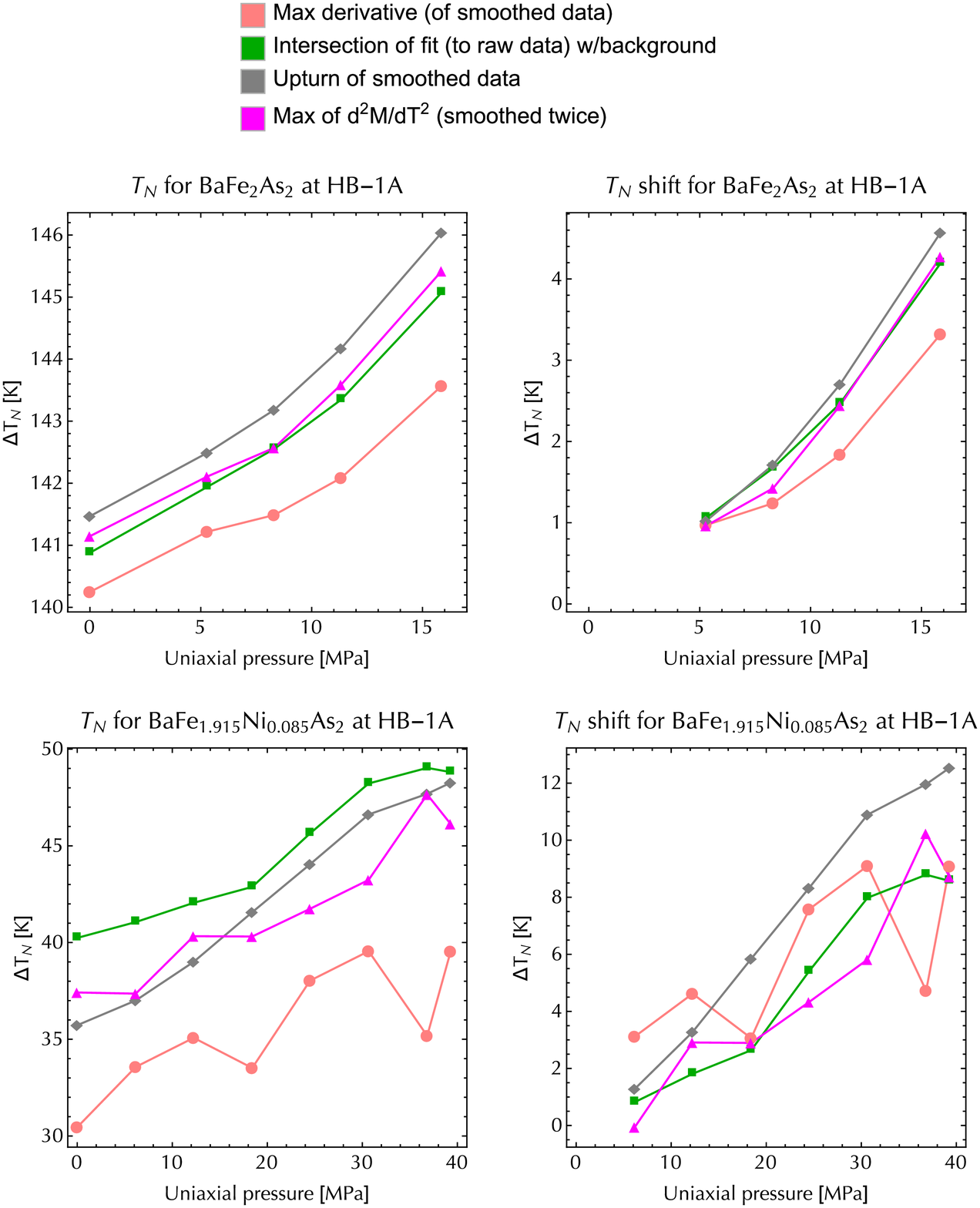}
\caption{Methods of calculating $T_N$.}
\label{tn-shift}
\end{figure*}

For all temperature scans, $T_N$ was calculated at each value of uniaxial pressure using four different methods, and the vertical error bars in Fig. 1 represent the differences among the methods:

\begin{enumerate}
  \item The maximum slope $dM/dT$ of the scan, after filtering by a windowed moving-average.
  \item The $T$-axis intercept of a linear fit to a few points nearby the temperature that maximizes $dM/dT$.
  \item The temperature at which the intensity settles to twice its high-temperature background value.
  \item The temperature that maximizes the second derivative $d^2M/dT^2$.
\end{enumerate}

The differences in $T_N$ between the scans at different pressures are compared with the zero pressure scan, and the derived values for $T_N$ represents the mean of these methods. The variability between the methods causes our estimates of the changes in $T_N$ to be far more precise ($\sim$0.5-1 K) than the values for $T_N$ themselves.

To explain precisely how we use this information, we show the results of the different methods in SFigure \ref{tn-shift}.  In Figure 1(d) in the main text, we show the differences in $T_N$ for BaFe$_2$As$_2$ rather than the calculated $T_N$ itself, in order to compare between the HB-1A and IN8 experiments, which used different samples with slightly different $T_N$ in the unstrained case. For this data, the error bars represent the difference between the extreme and the mean using the different $T_N$ methods. On the other hand, in Figure 1(f) in the main text we choose to show the actual values of $T_N$ calculated for BaFe$_{1.915}$Ni$_{0.085}$As$_2$. In this case, because the methods predict slightly different mean values as is evident in SFigure \ref{tn-shift}, calculating the error bars using the same method (extreme minus mean) leads to artificially large values. Instead, we use the standard deviation of the methods in Figure 1(f) of the main text.

\subsection{Pneumatic uniaxial pressure apparatus used at HB-1A and M20.}

\begin{figure}[h]
\includegraphics[scale=.43]{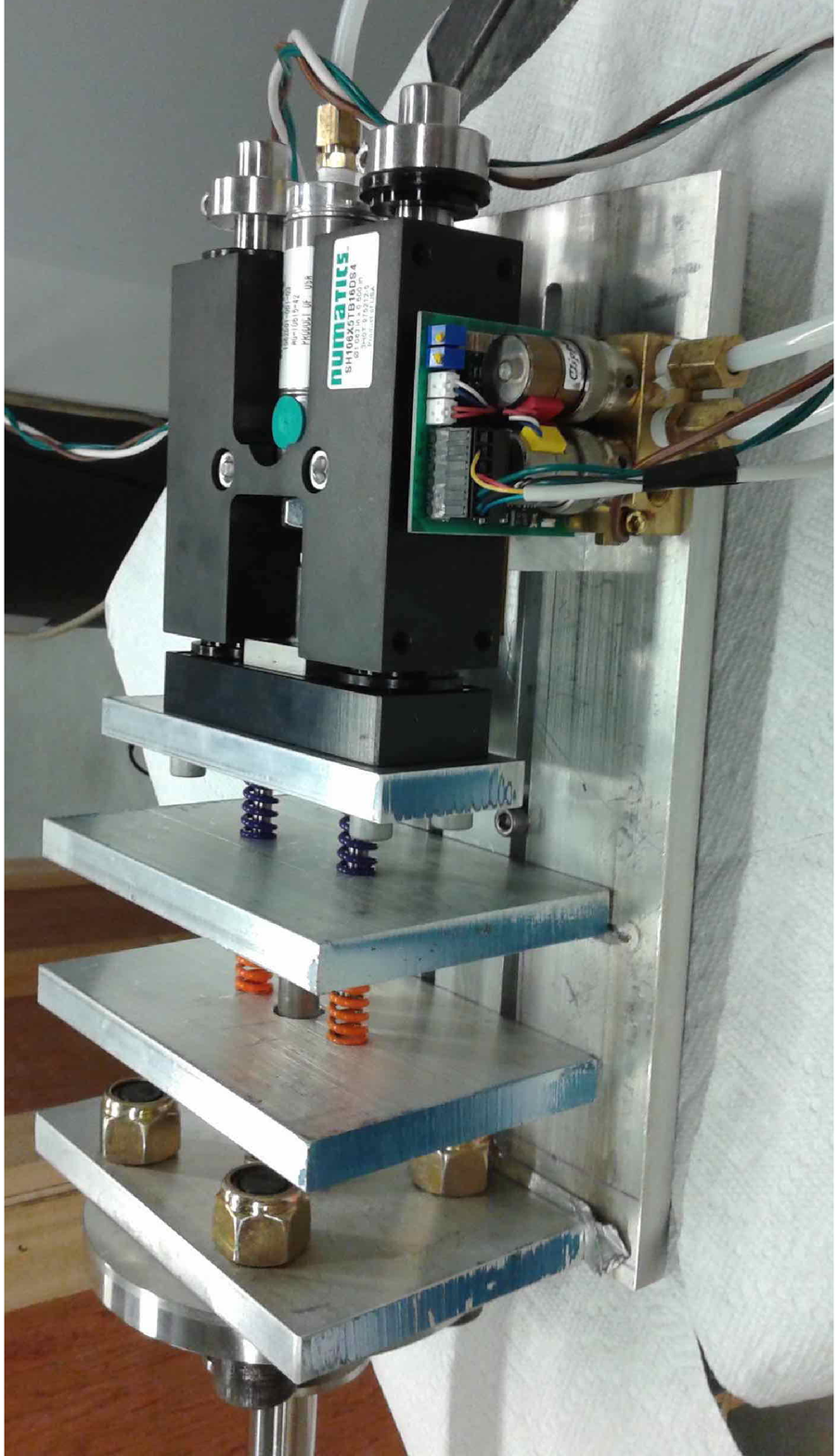}
\caption{Valve, thruster, and load-balancing components of the pneumatic uniaxial pressure apparatus.}
\label{pneumatic-head}
\end{figure}

\begin{figure}[h]
\includegraphics[scale=.4]{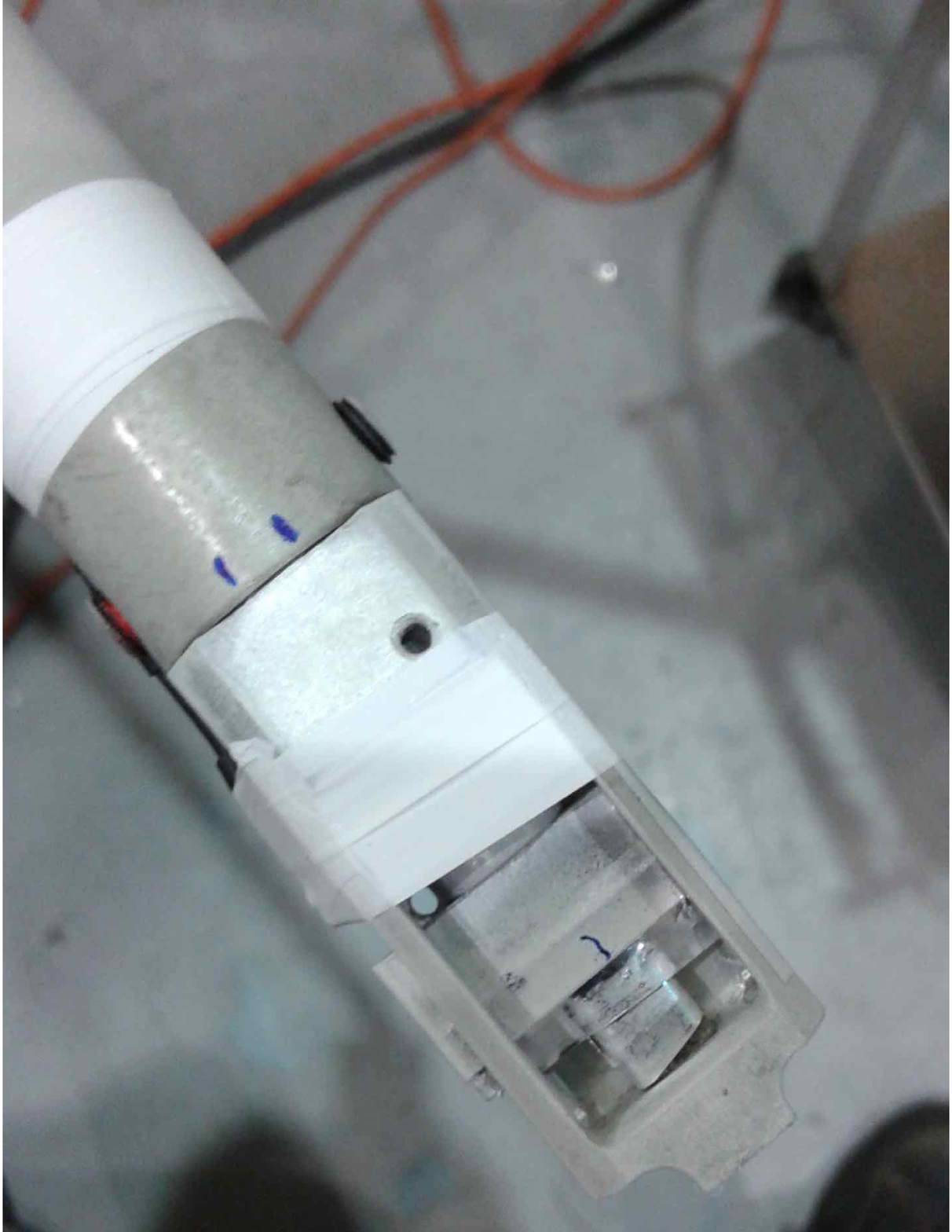}
\caption{Sample holder section of the pneumatic uniaxial pressure apparatus, shown with sample mounted at TRIUMF under thin aluminum foil tape.}
\label{pneumatic-sampleholder}
\end{figure}

\begin{figure*}[t]
\includegraphics[scale=.48]{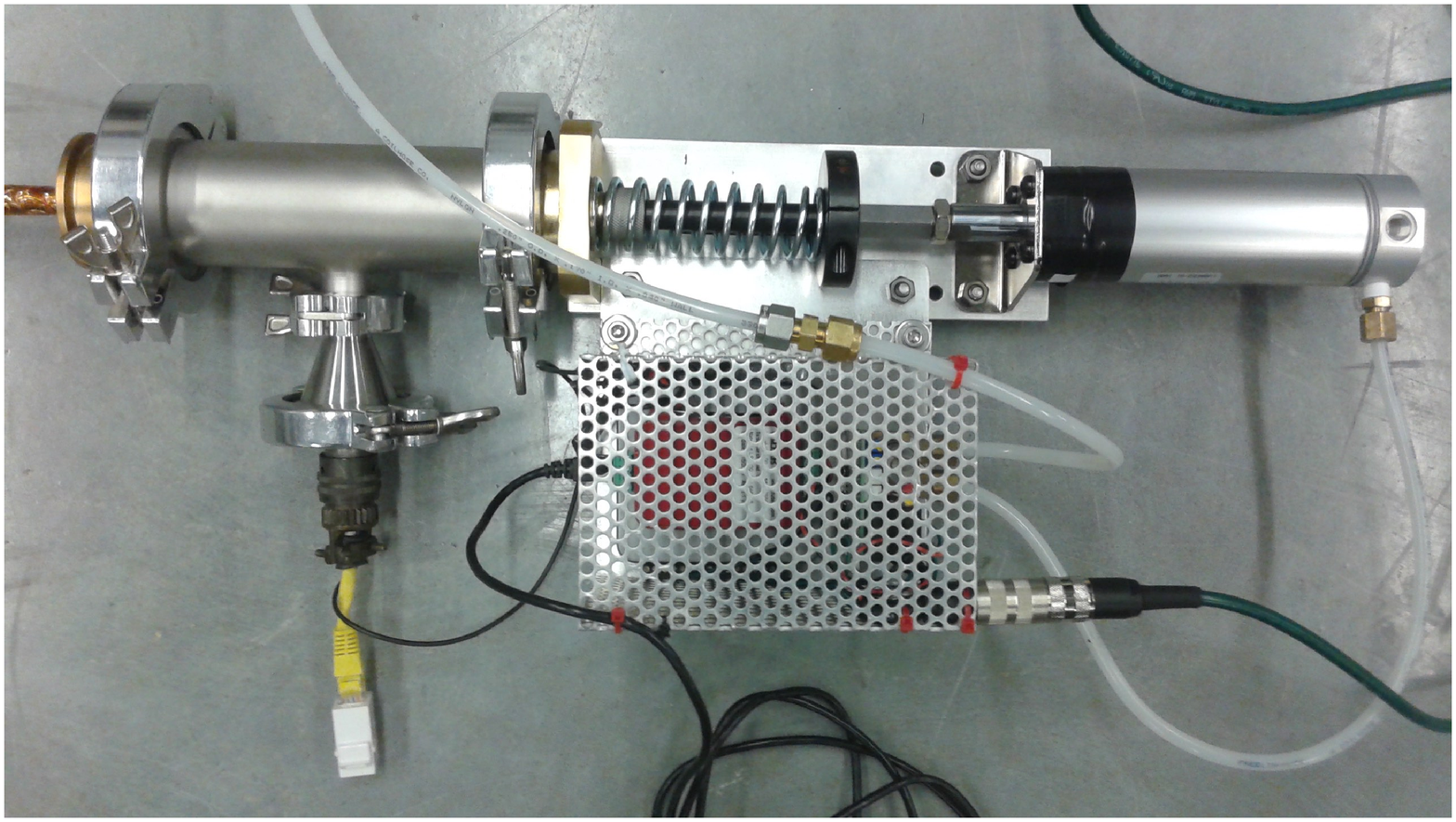}
\caption{Improved apparatus utilizing the secondary load cell.}
\label{pneumatic2-head}
\end{figure*}

In order to control uniaxial pressure with better precision than the spring-based clamps, and to ensure that the pressure remains constant regardless of thermal contraction of the sample and apparatus, we designed a uniaxial pressure device based on a pneumatic thruster in a feedback-controlled loop (SFigures \ref{pneumatic-head} and \ref{pneumatic-sampleholder}). Drawing air from a standard compressed gas cylinder, an electronically-controlled regulator (Proportion Air MPV1PBNEEZP150PSGAXL) was used to regulate the pressure in the reservoir of a linear thruster. The thruster piston was rigidly connected to a stainless steel tube passing inside an outer stainless steel tube, with both tubes running the length of the cryostat, and the inner tube pushing the sample against an all-aluminum frame held by the outer tube. The sample mount accommodate an adjustable position and pressure applied via the application of force from the linear thruster through the inner stainless steel tube. Outside the cryostat, one or more springs push against rigid plates attached to the inner stainless steel tube, in order to balance the weight of the device, to provide stiffness to overcome any friction in the system, and to provide flexibility in the mechanical linkage in order to keep the outputs stable and ensure good calibration even at low absolute pressures. A digital interface connected through a low-cost ADC allowed computer control of the regulator feedback loop setpoint over USB, which we integrated into the native software environment at both facilities.

Temperature control was achieved by mounting a sensor in thermal contact with the aluminum sample holder, and by mounting garolite baffles at several positions along the outer stainless steel tube. The temperature-induced differential length change along the length of both inner and outer stainless steel tubes is nominally identical, and therefore applies only small additional forces during temperature sweeps.

For the $\mu$SR experiments on the BaFe$_{1.9}$Co$_{0.1}$As$_2$ sample, we used an improved version of the apparatus (SFigure \ref{pneumatic2-head}) containing a load cell to measure and monitor the compression force in the inner stainless steel tube. The load cell is located inside the cryostat vacuum space and is therefore a direct probe of the force on the sample. We use the load cell output signal as the error signal for a second negative feedback loop, which maintains a constant thrusting force over long periods of time and wide temperature sweeps. The secondary feedback loop was controlled with a Labview interface. With the second loop closed, we monitored the actual regulator output and found that the pressure required to maintain constant force (as measured by the load cell) changes a small amount when the cryostat is at different temperatures.

The amount of aluminum around the sample was minimized in order to reduce the background, and in the case of $\mu$SR to provide a clear window for the muons to reach the sample. The crystal was free standing inside the aluminum sample holder, with no other external forces on the crystal. At M20, we added an additional mask over the sample holder to prevent muons from stopping in the 6061 aluminum, which contains some elemental impurities that could be magnetic; for the BaFe$_{1.915}$Ni$_{0.085}$As$_2$ crystal, we used a highly pure aluminum mask, and for the BaFe$_{1.9}$Co$_{0.1}$As$_2$ crystal a pure silver mask.

During the experiments, we stepped the pressure in steps of about 2-3 MPa, always reaching the setpoint from a lower pressure (except in the near-zero-pressure region), in order to overcome any mechanical hysteresis of the apparatus. In all experiments, the sample was warmed above $T_N$ before changing the pressure, then cooled to base temperature ($<$5 K), then warmed to the temperature under consideration. In the HB-1A experiments, the spectrometer was re-aligned to the nuclear peaks at base temperature before every measurement, although we only found only a tiny displacement of the sample between the zero- and finite-pressure warming cycles, with essentially no displacement as a function of pressure.

\subsection{Low temperature properties of BaFe$_{1.915}$Ni$_{0.085}$As$_2$.}

\begin{figure}[h]
\includegraphics[scale=.5]{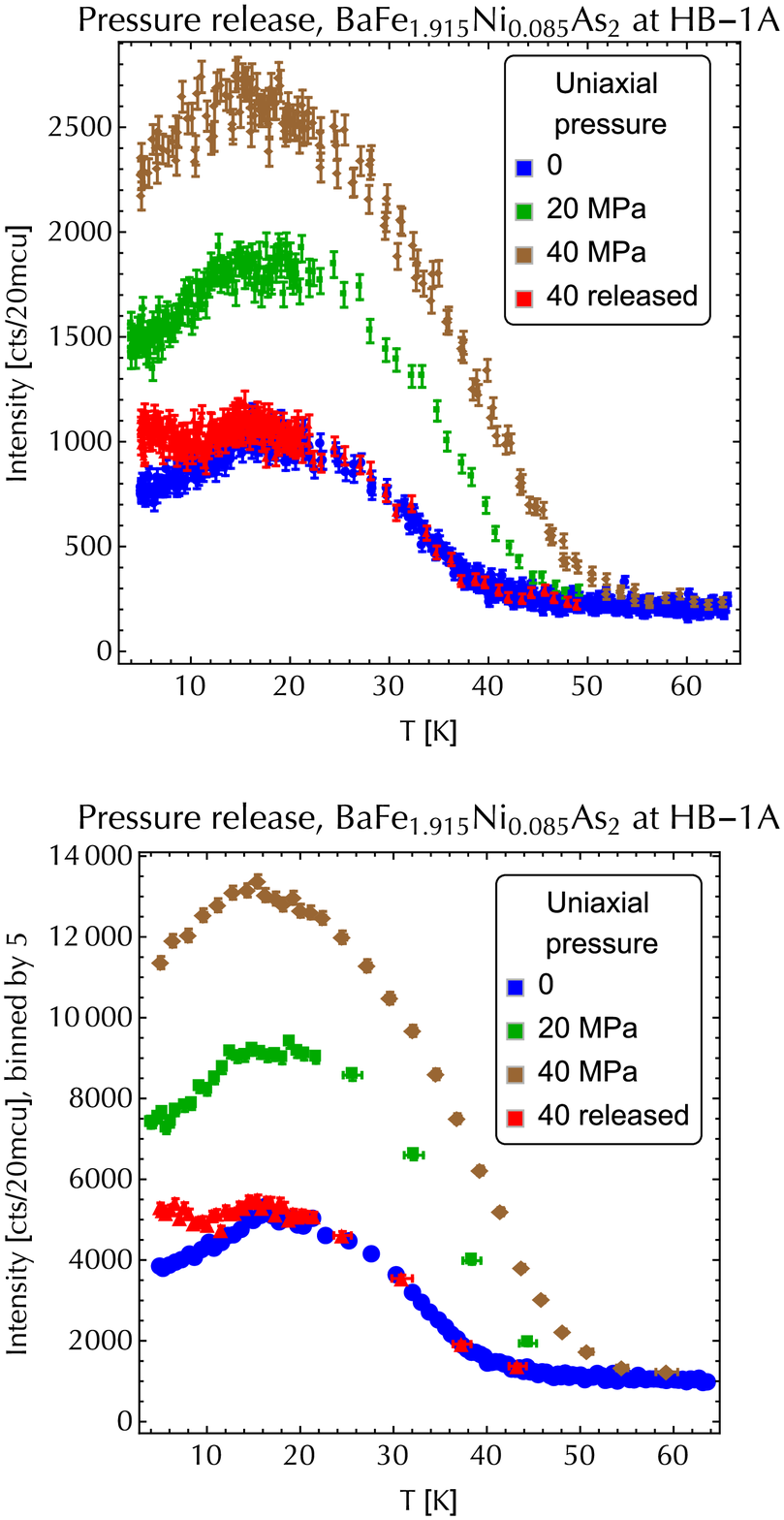}
\caption{Effect of releasing pressure at base temperature after cooling from $>$60 K under pressure.}
\label{pressurereleased}
\end{figure}

In SFigure \ref{pressurereleased}, we compare the temperature dependence of the strong Bragg position $\mathbf{Q}_\text{strong}$ in BaFe$_{1.915}$Ni$_{0.085}$As$_2$ at P = 0, 20, and 40 MPa, and with 40 MPa applied at room temperature and released at base temperature. The upper figure shows raw data and contains multiple temperature scans collected under each condition after thermal cycling above $\sim$ 60 K, which shows excellent agreement; for clarity, we combine neighboring points (groups of 5) and plot the same data in the lower figure. Comparing the 40 MPa data (brown) with the released data (red), it is clear the crystal does not relax into the fully twinned configuration (blue) \cite{2MLCZ15}. Upon warming, the crystal becomes fully twinned again near 15-20 K. The increased hardness of the material below $T_c$ has been previously observed in measurements of the shear modulus, and interpreted as evidence that the magnetism and lattice elasticity are entwined above $T_c$ \cite{2BoMe16,2FVBC10}. 
We also find that uniaxial pressure slightly decreases $T_c$ of BaFe$_{1.915}$Ni$_{0.085}$As$_2$.

\begin{figure}[h]
\includegraphics[scale=.55]{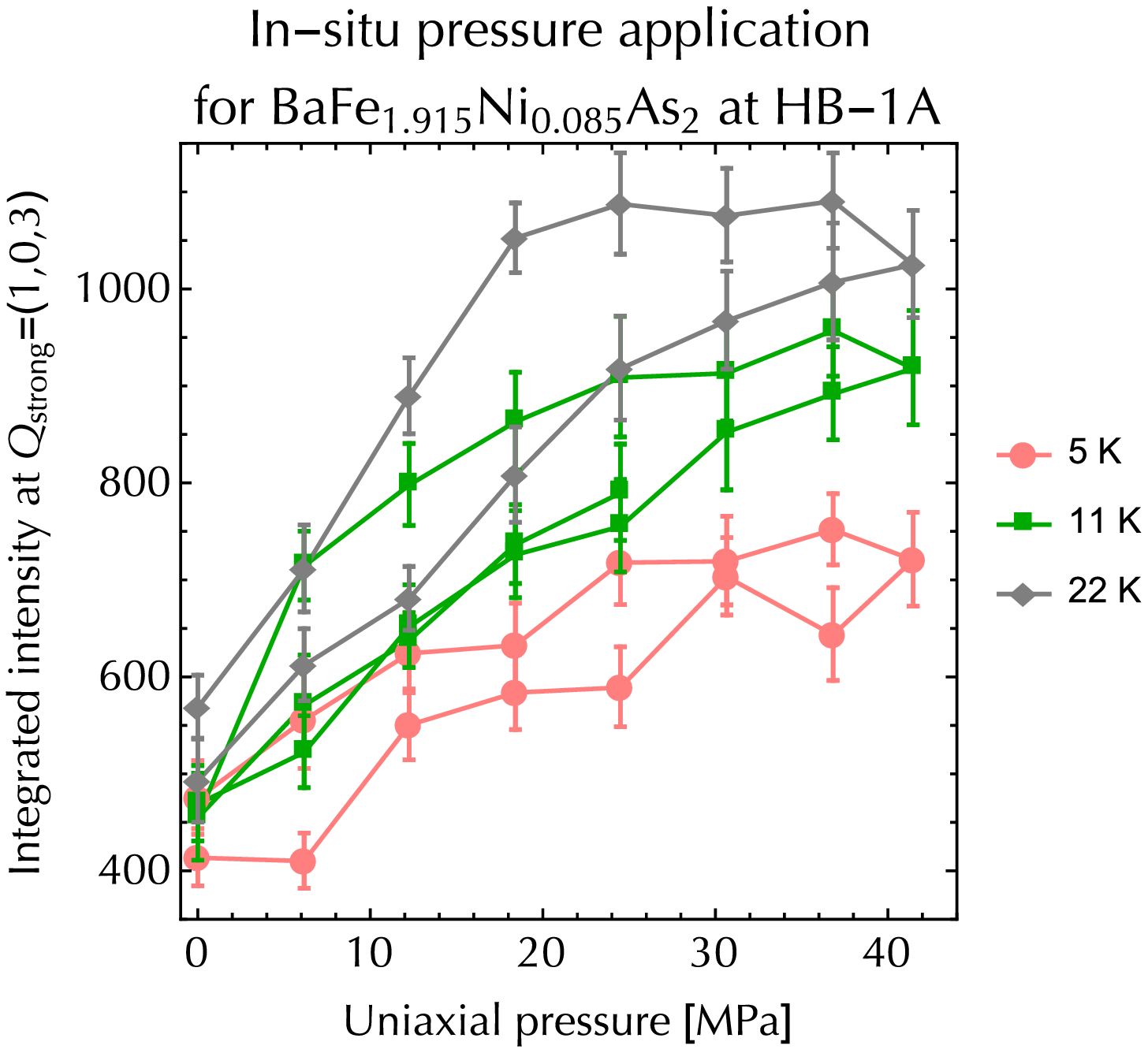}
\caption{Effect of uniaxial pressure controlled in-situ at low temperature. After cooling to base temperature in the pneumatic apparatus under ambient (unstrained) conditions, the sample was warmed to the temperature indicated and a rocking curve collected. Each point in the figure represents the integrated intensity of a Gaussian fit to the rocking curve.}
\label{ploopsplot}
\end{figure}

To understand the effect of flipping a single domain within the low-temperature region, we cooled the sample under unstrained conditions and then began applying pressure up to the maximum of $\sim$ 40 MPa, shown in SFigure \ref{ploopsplot}. We note a marked decrease in the effectiveness of uniaxial pressure in detwinning the crystal in this temperature range. This data also makes evident the necessity of warming in order to effectively apply uniaxial pressure.

\subsection{$\mu$SR experimental setup and raw data.}

\begin{figure}[h]
\includegraphics[scale=1.2]{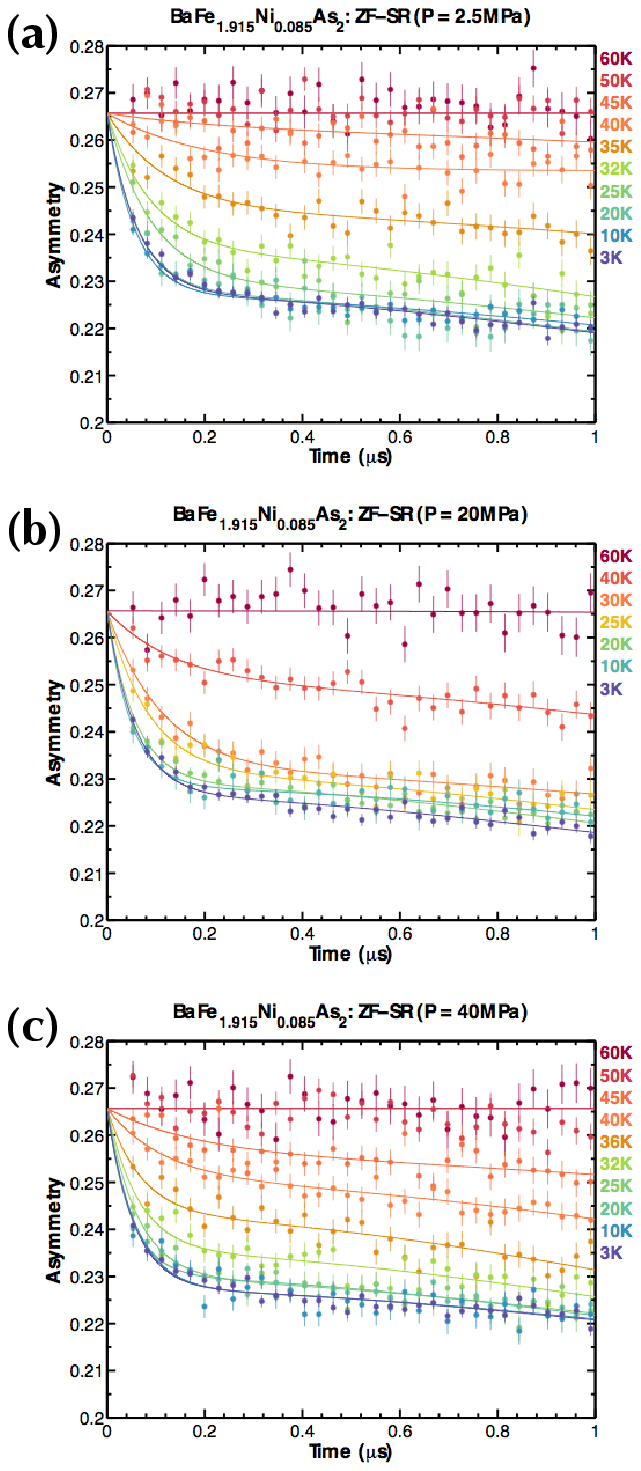}
\caption{Muon decay asymmetry for BaFe$_{1.915}$Ni$_{0.085}$As$_2$ as a function of time and temperature. The fast relaxing decay time is directly proportional to the ordered magnetic moment.
}
\label{fig-musr-ni}
\end{figure}

Since $\mu$SR can measure the magnetically ordered volume fraction and the ordered moment size independently from one another, it is ideally suited for assessing the detwinning effect on the magnetism of this material under uniaxial stress. An overview of the $\mu$SR technique can be found in \cite{2SoBK00}. At TRIUMF, 500 MeV protons collide with the surface of a light-nucleus target to produce pions, which quickly decay into (positive) muons that are completely spin-polarized opposite to the beam momentum. The muons are magnetically steered and focused onto the sample, where they dissipate energy until coming to rest within 1 ns at specific interstitial sites, without losing polarization. After a mean lifetime of 2.2 $\mu$s, the muons decay into positrons; due to maximal parity violation of the weak interaction, positrons are preferentially emitted in the direction of the muon spin, which has been precessing about the local magnetic field. The direction and magnitude of the internal field can then be probed by scintillators that detect the outgoing positron directions.
 
We rotated the muon spins in crossed electric and magnetic fields as they travel to the target, polarizing the spins to within 15 degrees of the downward-facing direction, which was along the sample $a$-axis (See Fig. 1(a) in the main text). The sample was mounted sideways in the OMNI-LAMPF spectrometer within a helium buffer gas cryostat, with uniaxial pressure applied horizontally along $b$, and $c$ facing into the beam. The sample was masked by placing a sheet of pure aluminum metal (pure silver for the BaFe$_{1.9}$Co$_{0.1}$As$_2$ crystal) over the entire sample holder, with a small opening to allow incident muons to impinge on the sample. The position of the cryostat was tuned by attaching a piece of nickel foil (which depolarizes the muons) over the mask window, then adjusting the cryostat position to minimize the paramagnetic asymmetry.

In SFigure \ref{fig-musr-ni}, we demonstrate the asymmetry versus decay time as a function of temperature fo the BaFe$_{1.915}$Ni$_{0.085}$ sample, at three values of uniaxial pressure studied (2.5, 20, and 40 MPa) in the zero-field (ZF) environment. The time spectra exhibit no oscillatory behavior and quickly relax at low temperatures within the first 0.5 microseconds. These observations indicate that the internal field is spatially disordered and inhomogeneous, in agreement with previous work on electron-doped BaFe$_{2}$As$_2$ \cite{2WABC14}. We modeled the ZF time spectra using a combination of a short-decay exponential and two Gaussian functions. Specifically, the ZF asymmetry time spectra were fit according to the model

\begin{equation}
\begin{aligned}
A_\text{ZF}(t) &= A_b e^{-\frac{1}{2}(\sigma_b t)^2} \\
&\quad + A_s \left[ (1 - F) e^{-\frac{1}{2}(\sigma_\text{para} t)^2} + F e^{-\lambda t} \right] 
\label{eqn:musr-ni}
\end{aligned}
\end{equation}

where $A(t)$ is the polarization function as a function of time; $A_b$ and $A_s$ are the asymmetry contributions from the background (sample holder) and sample, respectively; $\sigma_b$ and $\sigma_\text{para}$ are the Gaussian relaxation rates of the background and paramagnetic regions of the sample, respectively; $\lambda$ is the exponential relaxation rate for the magnetically ordered regions of the sample; and $F$ is the fractional contribution of the ordered phase to the sample asymmetry satisfying $0  \le  F \le 1$. Under this model of the ZF asymmetry spectra, the ordered moment size is proportional to the fast relaxation rate $\lambda$ and $F$ is proportional to the magnetically ordered volume fraction of the sample. This model was statistically refined against the experimental data using the musrfit software package with several parameters constrained globally, such as the fraction of the signal originating from the sample, and the total asymmetry.

The relaxation rates in the BaFe$_{1.915}$Ni$_{0.085}$ crystal appear to be peaked at $T\sim 20$ K, with the 40 MPa rate nearly 20\% greater than the relaxation rate at ambient pressure, suggesting that the magnetism is enhanced with increasing uniaxial strain. Despite the relatively low $\mu$SR statistics, the temperature and pressure dependence of the magnetic order parameter is consistent with the neutron scattering results. Additional $\mu$SR experiments were conducted with an external field applied longitudinally to the muon spin ($a$-axis), yielding time spectra that exhibit no critical spin dynamics.

\begin{figure*}[t]
\includegraphics[scale=0.4]{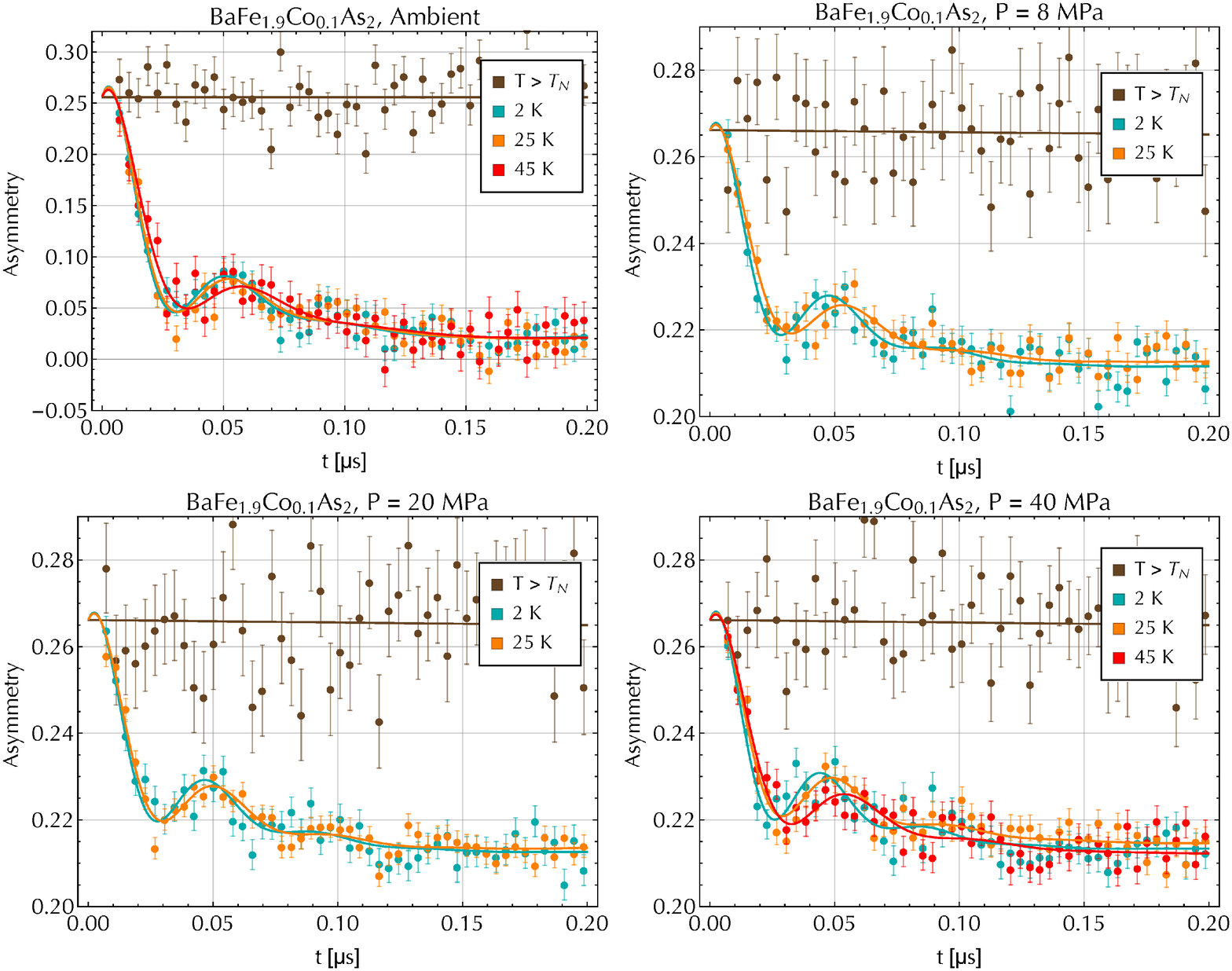}
\caption{Muon decay asymmetry for BaFe$_{1.9}$Co$_{0.1}$As$_2$ as a function of time and temperature. The fast relaxing oscillation frequency is directly proportional to the ordered magnetic moment.
}
\label{fig-musr-co-short}
\end{figure*}

\begin{figure*}[t]
\includegraphics[scale=0.4]{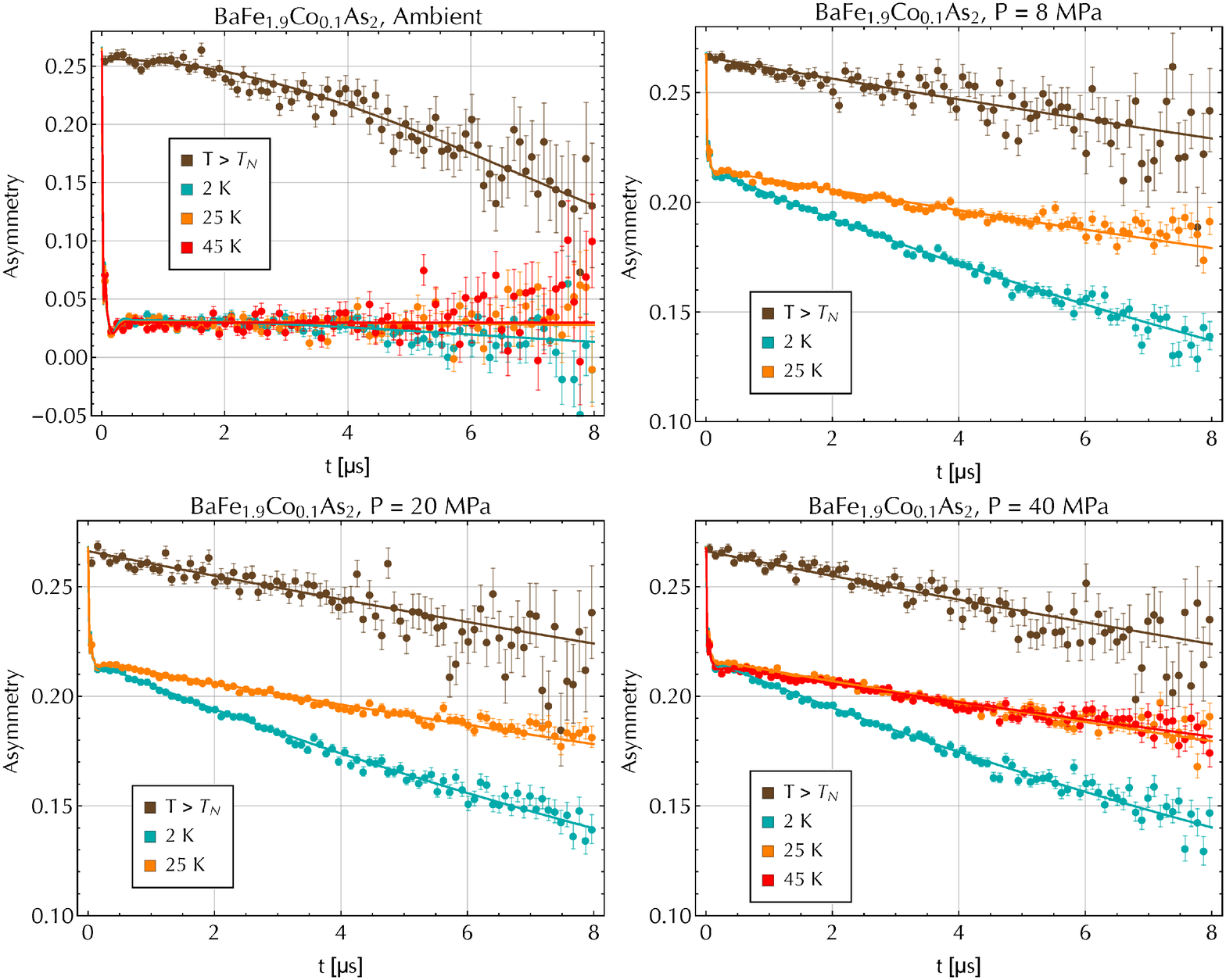}
\caption{Muon decay asymmetry for BaFe$_{1.9}$Co$_{0.1}$As$_2$ as a function of time and temperature.
}
\label{fig-musr-co-long}
\end{figure*}

In SFigures \ref{fig-musr-co-short} and \ref{fig-musr-co-long}, we demonstrate the asymmetry versus decay time as a function of temperature fo the BaFe$_{1.9}$Co$_{0.1}$As$_2$ sample in the zero-field (ZF) environment. As stated in the main text, we observed longer-lived oscillations in this sample that are more indicative of coherent long range order compared to the BaFe$_{1.915}$Ni$_{0.085}$ sample. We performed experiments at P=8, 20, and 40 MPa inside the uniaxial pressure instrument, and for the 0 pressure case the sample was transferred into the standard low-background sample holder where it is held in place with mylar tape. In the instrument, we found about a 4:1 ratio of muons stopping in the silver mask versus those stopping in the sample. Thus, at high temperature the muons landing in the paramagnetic sample are essentially impossible to detect underneath this background, and we have effectively absorbed its contribution as a small correction to the background term. We find the background function to then be different in this experiment compared to the BaFe$_{1.915}$Ni$_{0.085}$ experiment, that we attribute to the use of a silver mask instead an aluminum mask, which may support a different paramagnetic response due to differing levels of impurities, etc. In the ambient pressure case with the mylar tape, the situation is reversed: a Gaussian form describes the paramagnetic relaxation in the BaFe$_{1.9}$Co$_{0.1}$As$_2$ crystal, as it did in the in the BaFe$_{1.915}$Ni$_{0.085}$ experiment, while a very small background signal was observed, which we absorb as a small correction to the paramagnetic relaxation term. The overall model functions are given by

\begin{equation}
\begin{aligned}
A_\text{ZF,P=0}(t) =  A_\text{para} &e^{-\frac{1}{2}(\sigma_\text{para} t)^2} \\
\quad + (1-A_\text{para}) &\left[ F_1 e^{-\lambda_1 t} \text{cos}(\omega_1 t+ \phi_1)/\text{cos}(\phi_1) \right. \\
&\left. + (1-F_1) e^{-\lambda_2 t} \text{cos}(\omega_2 t+ \phi_2)/\text{cos}(\phi_2)  \right] \\
A_\text{ZF,P>0}(t) =  A_\text{b} &e^{-\lambda_\text{b}t} \\
\quad + (1-A_\text{b}) &\left[ F_1 e^{-\lambda_1 t} \text{cos}(\omega_1 t+ \phi_1)/\text{cos}(\phi_1) \right. \\
&\left. + (1-F_1) e^{-\lambda_2 t} \text{cos}(\omega_2 t+ \phi_2)/\text{cos}(\phi_2)  \right] \\
\label{eqn:musr-co}
\end{aligned}
\end{equation}

where $A(t)$ is the polarization function as a function of time; $A_\text{para}$ and $A_b$ are the asymmetry contributions from the background/paramagnetic regions for the cases described above, respectively; $\sigma_\text{para}$ and $\lambda_\text{b}$ are the relaxation rates of the background/paramagnetic regions; $F_1$ is the fraction of muons stopping at site 1, satisfying $0  \le  F_1 \le 1$, which we define as the site with the low ($\sim$2 MHz) oscillation frequency; $\lambda_1$ and $\lambda_2$ are the relaxation rates, $\omega_1$ and $\omega_2$ are the frequencies, and $\phi_1$ and $\phi_2$ are the phases for the two oscillatory components. Under this model of the ZF asymmetry spectra, the ordered moment size is proportional to the faster cosine frequency $\omega_2$. After refining all parameters separately except the total asymmetry $A_\text{para}$ or $A_\text{para}$, we found small changes in every parameter and therefore globally constrained all parameters (except $omega_1$) separately for P=0 and for P$>$0. We find small differences between these global parameters for the cases with and without pressure.

\subsection{Error bars in main text for neutron and $\mu$SR measurements}

In Fig. 1(g) of the main text, we determine the increase in magnetic moment for each sample based on linear fits to the data points.

For the $\mu$SR experiments on BaFe$_{1.9}$Co$_{0.1}$As$_2$, we show the fast cosine frequency data in Fig. 4(a) in the main text. For the point in Fig. 1(g), we choose the data at T = 25 K, the lowest measured temperature satisfying $T>T_\text{supercond.}$. The linear fit has standard errors for slope $\delta (\Delta M/\Delta P)$ and intercept $\delta M_\text{P=0}$, and the error bar is calculated by directly combining the errors, $\delta M_\text{P=0} + \delta (\Delta M/\Delta P)*40\text{ MPa}$.

For the $\mu$SR experiments on BaFe$_{1.915}$Ni$_{0.085}$As$_2$, we fit the fast relaxation rates at 0, 20, and 40 MPa, at the five lowest temperatures shown in Fig. 3(g) in the main text, which are in the magnetically saturated region. The error for each temperature is again calculated by directly combining the standard errors, $\delta M_\text{P=0} + \delta (\Delta M/\Delta P)*40\text{ MPa}$, and the result is averaged to form the error bar in Fig. 1(g).

For the neutron scattering experiments on BaFe$_{1.915}$Ni$_{0.085}$As$_2$, the point was determined from the integral of the rocking scan at maximum pressure (40 MPa) at $T = 20$ K. The ratio of the integral to the ambient pressure case is 2.528, corresponding to an increase in $M^2$ by 2.528/2 = 26.4 \%, and an increase in $M$ of 12.4 \%. The integral error of 0.115 was determined by fitting the scan to a Gaussian peak plus linear background, then letting the fit parameters vary over the range of their respective standard errors and choosing the global maximum, which is an overestimate. Using this result, the error bar is given by the ratio 0.115/2 = 0.057. To account for the overestimate in the integral method, and to account for the added certainty from the two independent measurements near 40 MPa that have nearly the same result, we have reduced this error bar by another factor of 2.

In Fig. 1(h) in the main text, we roughly estimate the magnetic ordering temperature $T_N$ for each sample and its error $\delta T_N$, and the shift under uniaxial pressure $\Delta T_N$ for each sample and its error $\delta \Delta T_N$. The error bars are determined by combining the errors $\frac{\Delta T_N}{T_N + \delta T_N} - \frac{\Delta T_N}{T_N}$ and $\frac{\Delta T_N + \delta \Delta T_N}{T_N} - \frac{\Delta T_N}{T_N}$ in quadrature.

\subsection{Details of Density Functional Theory + Random Phase Approximation calculations.}

To better understand the enhancement of the magnetic instability under uniaxial pressure in BaFe$_2$As$_2$ and BaFe$_{1.915}$Ni$_{0.085}$As$_2$, we study the magnetic susceptibility in the normal (unstrained) and uniaxially-strained paramagnetic phase within a random-phase approximation (RPA) approach. Several earlier works have presented microscopic models that can account for the increase in $T_N$, but they have relied on pure phenomenology \cite{2CCEP10,2KACF12} or on effective Hamiltonians based on local moment interactions \cite{2ACano12,2HuSK12,2QDLR15,2TJFV13}. In one case, uniaxial pressure was found to reduce the ordered moment of BaFe$_2$As$_2$ \cite{2TJFV13}. Our calculations consider only the itinerant degrees of freedom, and we show that uniaxial pressure enhances the spin susceptibility at the AF wave vector, thus driving an upward shift in $T_N$ for both BaFe$_2$As$_2$ and BaFe$_{1.915}$Ni$_{0.085}$As$_2$. We find that the enhancement is mainly a consequence of better Fermi surface nesting under uniaxial pressure, while a slight redistribution of orbital weights on the Fermi surface is responsible for symmetry breaking between $(\pm 1,0)$ and $(0,\pm 1)$ that creates a preferred orientation for the formation of AF order. This result suggests that the onset of AF order and the low-temperature magnetic properties are deeply connected to the dynamics of the itinerant electrons.

\begin{figure}[h]
\includegraphics[scale=0.32]{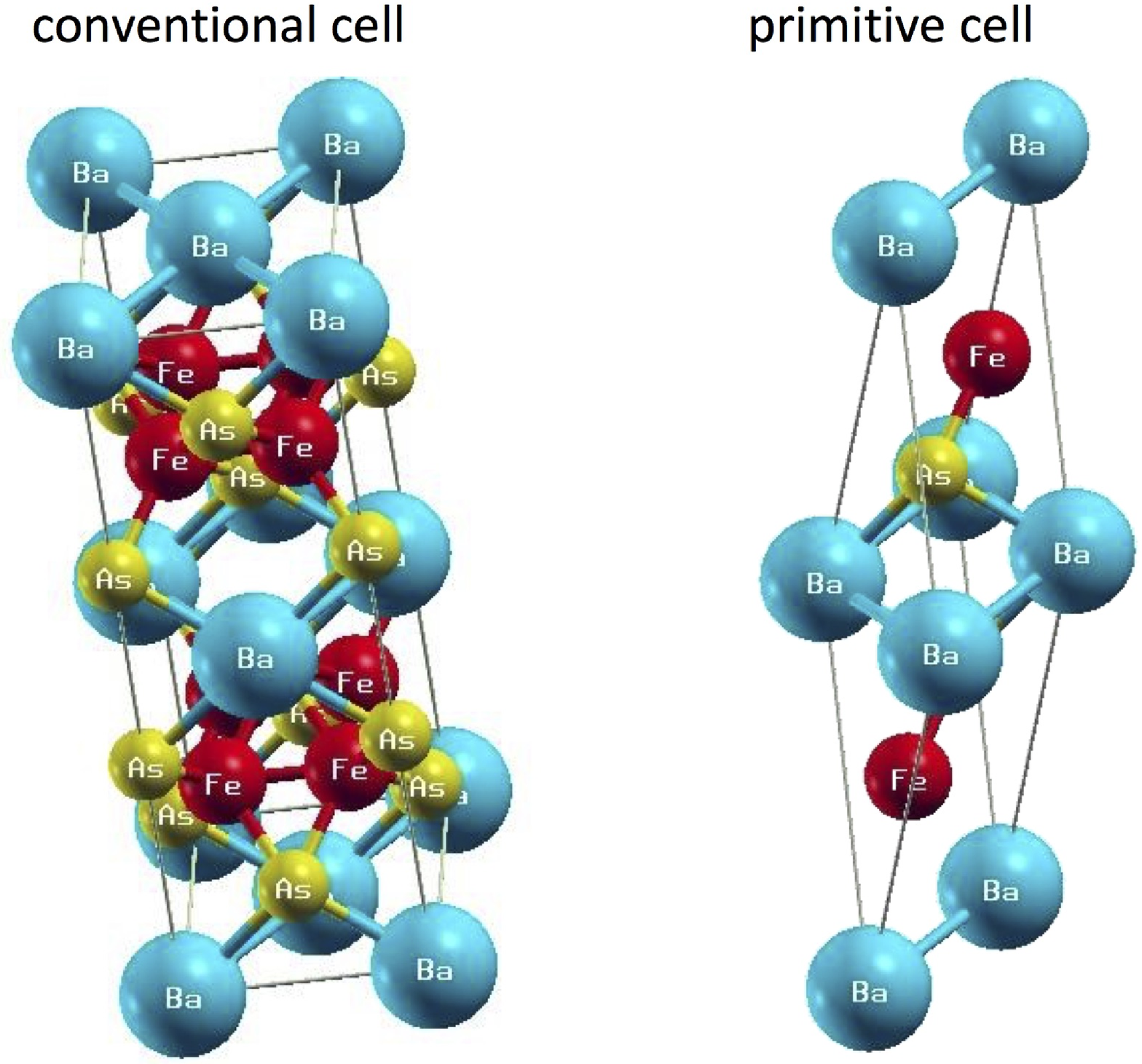}
\caption{Unit cells employed in the DFT calculations.
}
\label{DFT}
\end{figure}

\begin{figure}[h]
\includegraphics[scale=.4]{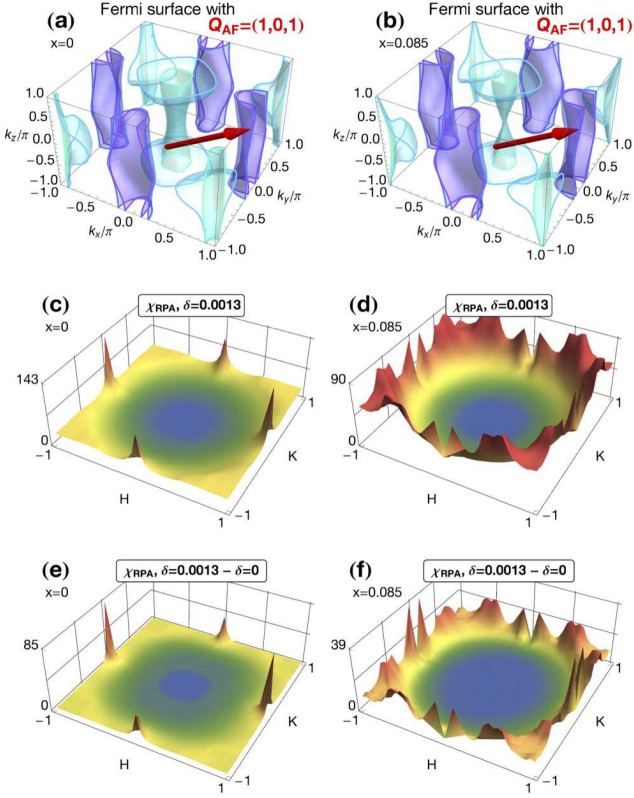}
\caption{  (Color online) 
(a-b) Three-dimensional Fermi surfaces in the one-iron Brillouin zone for BaFe$_2$As$_2$ and BaFe$_{1.915}$Ni$_{0.085}$As$_2$.
(c) RPA susceptibility ($U=1.23$) for BaFe$_2$As$_2$ with orthorhombic lattice distortion $\delta=0.0013$, equivalent to 10 MPa.
(d) RPA susceptibility ($U=1.32$) for BaFe$_{1.915}$Ni$_{0.085}$As$_2$ with $\delta =0.0013$, equivalent to roughly 30-40 MPa in this material. For the same Coulomb interaction parameters used in BaFe$_2$As$_2$ ($U=1.23$), the AF peaks are much less pronounced.
(e) Figure (c) for BaFe$_2$As$_2$ with the unstrained ($P=0$) result subtracted, demonstrating enhancement at 
$\mathbf{Q}_\text{strong}=(\pm 1,0)$ and a more modest enhancement at $\mathbf{Q}_\text{weak}=(0,\pm1)$.
(f) Similar results for BaFe$_{1.915}$Ni$_{0.085}$As$_2$.
}
\end{figure}

We start by generating tight-binding Hamiltonians using DFT and the Wannier function technique, for the parent compound BaFe$_2$As$_2$ with and without uniaxial distortion. First, we scale the room-temperature lattice parameters for BaFe$_2$As$_2$ ($a_\text{RT}=b_\text{RT}=5.60$ \AA\ and $c_\text{RT}=13.02$ \AA), to 140 K by interpolating the results of differential capacitance dilatometry measurements \cite{2BNNS09}, finding $a_0=b_0=5.596$ \AA, $c_0=12.95$ \AA\  in the stress-free case. To obtain the distortion under 10 MPa, we estimate the lattice asymmetry, $\delta_{ab} \equiv (a_0-b_0)/(a_0+b_0) \approx 0.0013$, from high-precision experimental data on uniaxially-strained BaFe$_{1.97}$Ni$_{0.03}$As$_2$ at its $T_N$ = 120 K \cite{2XYLu15}. We assume is that the distortion in this case is the same for the parent compound at $T_N$ =140 K, and that uniaxial pressure preserves the overall unit cell volume, arriving at $a=a_0(1+\delta)=5.603$ \AA, $b=a_0(1-\delta)=5.589$ \AA, with $c$ remaining 12.95 \AA. This distortion is much smaller than the natural orthorhombic in the low-temperature state ($\delta_{ab} \approx 0.075$) \cite{2WaDN09}, and can therefore be viewed as a perturbation. (We note that due to the bilinear coupling of the orthorhombic distortion and the electronic nematic degree of freedom, i.e. simultaneous onset, a small strain field can enhance and even spontaneously generate nematic order \cite{2Paul14}.)

The Fermi surface found using these models [SFig. 8(a) and 8(b)] is similar to that reported in previous calculations \cite{2Sing08}. To conduct calculations for BaFe$_{1.915}$Ni$_{0.085}$As$_2$, we perform a rigid band shift; in this compound, the bulk modulus is slightly harder and the distortion is equivalent to 30-40 MPa (see main text). The hollow arrows in Figures 1(b) and 1(c) in the main text show the direction of pressure-induced changes in the Fermi surfaces for BaFe$_2$As$_2$ and BaFe$_{1.915}$Ni$_{0.085}$As$_2$, respectively, while in SFigures 8(a) and 8(b) we show the three-dimensional Fermi surface for each compound in the one-iron Brillouin zone, with one position of the AF nesting wave vector marked in red.

The Density Functional Theory (DFT) calculations are perfomed with WIEN2K \cite{2wien2k} and the Wannier transformation is performed with WIEN2WANNIER \cite{2wien2wannier} and WANNIER90 \cite{2wannier90}.The conventional and primitive unit cell are depicted in SFigure \ref{DFT} . The number of LAPW basis functions was set by RKmax=7 and the k-mesh was taken to be 7x7x7. (For the Wannier transformation a k-mesh of 12x12x12 was used). Atomic relaxations were converged down to 2 mRy/bohr. For the Wannier transformation, the 10 Fe-3d orbitals were projected on bands. In the Wannier90 input file we used num$\textunderscore$iter=0, dis$\textunderscore$win$\textunderscore$min=-2.4, dis$\textunderscore$win$\textunderscore$max=3.2 and dis$\textunderscore$num$\textunderscore$iter=100.

Our calculation of the static RPA spin susceptibility is identical to that described in our previous work \cite{2MaSc08,2MGSH09,2MaHS12}. We calculate the orbital- and momentum-resolved noninteracting dielectric function \cite{2Miha11,2WKZB13,2KOAU08},

\begin{widetext}
\begin{equation}
\begin{aligned}
\chi^0_{\ell_1\ell_2\ell_3\ell_4} (\mathbf{q},\omega=0) = - \frac{1}{N_\text{Fe}} \frac{1}{N_k} \sum_{\mathbf{k},\mu\nu} \
\frac{a^{\ell_4}_\mu(\mathbf{k}) a^{\ell_2,*}_\mu(\mathbf{k}) a^{\ell_1}_\nu(\mathbf{k}+\mathbf{q}) a^{\ell_3,*}_\nu(\mathbf{k}+\mathbf{q})}
{E_\mu(\mathbf{k}) - E_\nu(\mathbf{k}+\mathbf{q}) + i\delta} \
\left( f[E_\mu(\mathbf{k}),kT] - f[E_\nu(\mathbf{k}+\mathbf{q}),kT] \right) \qquad ,
\label{eqn:chi0}
\end{aligned}
\end{equation}
\end{widetext}

with $N_\text{Fe}$=2 the number of iron sites per unit cell, $N_k$ the number of grid points in momentum ($k$) space, band indices $\mu$ and $\nu$, orbital indices $\ell$. The matrix elements are represented by the orbital projection of the Bloch state, $a^\ell_\mu = \langle \ell|\mu\mathbf{k} \rangle$, $f[E,kT]$ is the Fermi function at temperature $T$, and the small parameter $\delta=0.01$ enforces analyticity in the sum over Matsubara frequencies. We sum over a k-space mesh of between 100x100x8 and 160x160x32 points over the 3D Brillouin zone depending on the region of interest. We then account for the interactions $\mathcal{U}$ (containing same- and different- orbital Coulomb repulsion $U$ and $U'$  as well as Hund's rule and pair-hopping couplings $J$ and $J'$) using the usual RPA form for the susceptibility, $\chi_\text{RPA}=\chi_0/(1-\mathcal{U}\chi_0)$, while constraining $U'=U/2$, $J=J'=U/4$ consistent with local spin rotational invariance \cite{2NoRo14}. All calculations are done in the paramagnetic state and use an odd integer for $L$ (q$_z$) enabling a direct comparison with neutron scattering. We choose $U=1.23$ eV for BaFe$_2$As$_2$ and $U=1.32$ eV for BaFe$_{1.915}$Ni$_{0.085}$As$_2$, just below the divergence of $\chi_\text{RPA}$ over the full Brillouin zone in both cases. We generate models with and without lattice distortion for the parent compound using $\delta_{ab} \equiv (a-b)/(a+b) \approx 0.0013$ representing the effect of uniaxial pressure. The $x=0.085$ case is considered by applying a rigid band shift (about 58 meV for both the strained and unstrained models), and we perform calculations at the experimentally observed $T_N^\text{(x=0)}=140 \text{ K}$ and $T_N^\text{(x=0.085)}=40 \text{ K}$.

For BaFe$_2$As$_2$, we find the intensity at the $\mathbf{Q}_\text{strong}$ magnetic position is enhanced by $\sim$100\% in the model including lattice distortion relative to the zero-pressure (ambient) case, while at the weak position it is also enhanced but only by $\sim$50\%. For BaFe$_{1.915}$Ni$_{0.085}$As$_2$, the instability is significantly less pronounced even for stronger RPA interactions, and the susceptibility peaks are incommensurate, consistent with previous calculations \cite{2LTZL14}; nevertheless, the trend is consistent with the BaFe$_2$As$_2$ results, demonstrating an overall increase and symmetry-breaking favoring $\mathbf{Q}_\text{strong}$ [SFig. 8(c) and 8(d)]. The differences under pressure are shown in SFigures 8(e) and 8(f).

To better understand the role of Fermi surface nesting alone, we suppress the orbital matrix elements in the bare susceptibility, cf. Eq. \ref{eqn:chi0} (we note that the distortion is very small, and rather different for the various hole and electron pockets in different $k_z$ planes, essentially prohibiting an analysis by inspection):

\begin{widetext}
\begin{equation}
\begin{aligned}
\chi^0_\text{FS} (\mathbf{q},\omega=0) = - \frac{1}{N_\text{Fe}} \frac{1}{N_k} \sum_{\mathbf{k},\mu\nu} \
\frac{1}
{E_\mu(\mathbf{k}) - E_\nu(\mathbf{k}+\mathbf{q}) + i\delta} \
\left( f[E_\mu(\mathbf{k}),kT] - f[E_\nu(\mathbf{k}+\mathbf{q}),kT] \right) \qquad .
\label{eqn:chi0noorb}
\end{aligned}
\end{equation}
\end{widetext}

Using this expression, we find a roughly equal enhancement of $\sim$1\% at both strong and weak Bragg positions, whereas with the matrix elements included, we find the symmetry breaking that is further enhanced by the RPA interactions. The results are summarized in Table \ref{tab:rpa}.

\begin{table}[h!]
\footnotesize
  \begin{center}
    \caption{Peak values of bare and RPA susceptibility.}
    \label{tab:rpa}
    \begin{tabular}{l|l|l|l|l}
       & AFM peak & Without $a(\mathbf{k})$ & With $a(\mathbf{k})$ & $\chi_\text{RPA}$\\
      \hline
      \hline
      x=0 & & & & (U=1.23) \\
      & P=0 & 1.0063 & 1.8205 & 58.3 \\
      & $\mathbf{Q}_\text{strong}$ & 1.0155 (+0.9\%) & 1.8464 (+1.4\%) & 144 (+147\%) \\
      & $\mathbf{Q}_\text{weak}$ & 1.0151 (+0.9\%) & 1.8403 (+1.1\%) & 94.3 (+61\%) \\
      \hline
      x=0.085 & & & & (U=1.32) \\
      & P=0 & 0.9563 & 1.7174 & 55 \\
      & $\mathbf{Q}_\text{strong}$ & 0.9642 (+0.8\%) & 1.7362 (+1.1\%) & 95 (+73\%)  \\
      & $\mathbf{Q}_\text{weak}$ & 0.9638 (+0.8\%) & 1.7332 (+0.9\%) & 78 (+42\%) \\
    \end{tabular}
  \end{center}
\end{table}

With this information, we conclude that the increase in $T_N$ in both compounds is mostly due to an enhancement of the Fermi surface nesting condition under uniaxial pressure, while the changes in the orbital weights due to pressure account for the asymmetry between $\mathbf{Q}_\text{strong}$ and $\mathbf{Q}_\text{weak}$.


\begin{thebibliography}{}
\bibitem{kamihara} Y. Kamihara, T. Watanabe, M. Hirano, and H. Hosono, J. Am. Chem. Soc. \textbf{130}, 3296-3297 (2008).
\bibitem{cruz} C. de la Cruz \textit{et al.}, Nature (London) \textbf{453},899 (2008).
\bibitem{qhunag} Q. Huang, Y. Qiu, W. Bao, M. A. Green, J. W. Lynn, Y. C. Gasparovic, T. Wu, G. Wu, and X. H. Chen, Phys. Rev. Lett. {\bf 101}, 257003 (2008).
\bibitem{WYRF09} S. D. Wilson, Z. Yamani, C.R. Rotundu, B. Freelon, E. Bourret-Courchesne, and R.J. Birgeneau, Phys. Rev. B {\bf 79}, 184519 (2009).
\bibitem{mgkim} M. G. Kim, R. M. Fernandes, A. Kreyssig, J. W. Kim, A. Thaler, S. L. Bud'ko, P. C. Canfield, R. J. McQueeney, J. Schmalian, and A. I. Goldman, Phys. Rev. B {\bf 83}, 134522 (2011).
\bibitem{scalapino} D. J. Scalapino, Rev. Mod. Phys. {\bf 84}, 1383 (2012).
\bibitem{dai} P. C. Dai, Rev. Mod. Phys. {\bf 87}, 855 (2015).
\bibitem{IIMazin} I. I. Mazin, M. D. Johannes, L. Boeri, K. Koepernik, and D. J. Singh, Phys. Rev. B {\bf 78}, 085104 (2008).
\bibitem{hirschfeld} P. J. Hirschfeld, M. M. Korshunov, I. I. Mazin, Rep. Prog. Phys. {\bf 74}, 124508 (2011).
\bibitem{chubukov} A. Chubukov, Ann. Rev. Condens. Matter Phys. {\bf 3}, 57 (2012).
\bibitem{supplementary} For additional data and analysis, see supplementary information. 
\bibitem{si} Q. Si and E. Abrahams, Phys. Rev. Lett. {\bf 101}, 076401 (2008).
\bibitem{cfang} C. Fang, H. Yao, W. F. Tsai, J. P. Hu, and S. A. Kivelson, Phys. Rev. B {\bf 77}, 224509 (2008).
\bibitem{xuc} C. K. Xu, M. M$\rm \ddot{u}$ller, and S. Sachdev, Phys. Rev. B {\bf 78}, 020501(R) (2008).
\bibitem{Basov11} D. N. Basov and A. V. Chubukov, Nat. Phys. {\bf 7}, 272 (2011).
\bibitem{Avci13} S. Avci, O. Chmaissem, J. M. Allred,	S. Rosenkranz, I. Eremin,	A. V. Chubukov,	D. E. Bugaris, D. Y. Chung,	M.G. Kanatzidis, J.-P Castellan, J. A. Schlueter,	H. Claus,	D. D. Khalyavin, P. Manuel,	A. Daoud-Aladine, and R. Osborn, Nat. Comm. {\bf 5}, 3845 (2013).
\bibitem{Waber} F. Wa$\beta$er, A. Schneidewind, Y. Sidis, S. Wurmehl, S. Aswartham, B. Buchner, and M. Braden, Phys. Rev. B {\bf 91}, 060505(R) (2015).
\bibitem{Allred16} J. M. Allred, K. M. Taddei, D. E. Bugaris,	M. J. Krogstad, S. H. Lapidus, D. Y. Chung, H. Claus,	M. G. Kanatzidis,	D. E. Brown, J. Kang,	R. M. Fernandes, I. Eremin,	S. Rosenkranz, O. Chmaissem, and R. Osborn, Nature Physics {\bf 12}, 493 (2016).	
\bibitem{Pratt12} D. K. Pratt, M. G. Kim, A. Kreyssig, Y. B. Lee, G. S. Tucker, A. Thaler, W. Tian, J. L. Zarestky, S. L. Bud'ko, P. C. Canfield, B. N. Harmon, A. I. Goldman, and R. J. McQueeney, Phys. Rev. Lett. {\bf 106}, 257001 (2012). 
\bibitem{HQluo12} H. Q. Luo, R. Zhang, M. Laver, Z. Yamani, M. Wang, X. Y. Lu, M. Y. Wang, Y. Chen, S. L. Li, S. Chang, J. W. Lynn, 
and P. C. Dai, Phys. Rev. Lett. {\bf 108}, 247002 (2012).
\bibitem{Dioguardi13} A. P. Dioguardi, J. Crocker, A. C. Shockley, C. H. Lin, K. R. Shirer, D. M. Nisson, M. M. Lawson, N. apRoberts-Warren, P. C. Canfield, S. L. Bud'ko, S. Ran, and N. J. Curro, Phys. Rev. Lett. {\bf 111}, 207201 (2013).
\bibitem{Dioguardi} A. P. Dioguardi, M. M. Lawson, B. T. Bush, J. Crocker, K. R. Shirer, D. M. Nisson, T. Kissikov, S. Ran, S. L. Bud'ko, P. C. Canfield, S. Yuan, P. L. Kuhns, A. P. Reyes, H.-J. Grafe, and N. J. Curro, Phys. Rev. B {\bf 92}, 165116 (2015).
\bibitem{XYLu14a} X. Y. Lu, D. W. Tam, Chenglin Zhang, Huiqian Luo, Meng Wang, Rui Zhang, Leland W. Harriger, T. Keller, B. Keimer, L.-P. Regnault, Thomas A. Maier, and P. C. Dai, Phys. Rev. B {\bf 90}, 024509 (2014).
\bibitem{KCPK16} H.-H. Kuo, J.-H. Chu, J.C. Palmstrom, S.A. Kivelson, and I.R. Fisher, Science {\bf 352}, 958 (2016). 
\bibitem{FeSc12} R. M. Fernandes, A. V. Chubukov, and J. Schmalian, Nat. Phys. {\bf 10}, 97 (2014).
\bibitem{anna} A. E. B$\rm \ddot{o}$hmer and C. Meingast, Comptes Rendus Physique {\bf 17}, 90 (2016).
\bibitem{jhchu} J. H. Chu {\it et al.}, Science {\bf 329}, 824 (2010).
\bibitem{matanatar} M. A. Tanatar, E. C. Blomberg, A. Kreyssig, M. G. Kim, N. Ni, A. Thaler, S. L. Bud'ko, P. C. Canfield, A. I. Goldman, I. I. Mazin, and R. Prozorov, Phys. Rev. B {\bf 81}, 184508 (2010).
\bibitem{fisher} I. R. Fisher, L. Degiorgi, and Z. X. Shen, Rep. Prog. Phys. {\bf 74}, 124506 (2011).
\bibitem{dhital} C. Dhital, Z. Yamani, W. Tian, J. Zeretsky, A. S. Sefat, Z. Wang, R. J. Birgeneau, and S. D. Wilson, Phys. Rev. Lett. {\bf 108}, 087001 (2012).
\bibitem{Dhital14} C. Dhital, T. Hogan, Z. Yamani, R. J. Birgeneau, W. Tian, M. Matsuda, A. S. Sefat, Z. Wang, and S. D. Wilson, Phys. Rev. B {\bf 89}, 214404 (2014). 
\bibitem{YSong13} Y. Song, S. V. Carr, X. Y. Lu, C. L. Zhang, Z. C. Sims, N. F. Luttrell, S. X. Chi, Y. Zhao, J. W. Lynn, and P. C. Dai, Phys. Rev. B {\bf 87}, 184511 (2013). 
\bibitem{xylu14} X. Y. Lu, J. T. Park, R. Zhang, H. Q. Luo, A. H. Nevidomskyy, Q. Si, and P. C. Dai, Science {\bf 345}, 657 (2014).
\bibitem{man} H. R. Man, X. Y. Lu, J. S. Chen, R. Zhang, W. L. Zhang, H. Q. Luo, J. Kulda, A. Ivanov, T. Keller, E. Morosan, Q. Si, and P. C. Dai, Phys. Rev. B {\bf 92}, 134521 (2015).  
\bibitem{XYLu15} X. Y. Lu, K. F. Tseng, T. Keller, W. L. Zhang, D. Hu, Y. Song, H. R. Man, J. T. Park, H. Q. Luo, S. L. Li, A. N. Nevidomskyy, and P. C. Dai, Phys. Rev. B {\bf 93}, 134519 (2016).
\bibitem{Tomic} M. Tomic, H. O. Jeschke, R. M. Fernandes, and R. Valenti, Phys. Rev. B {\bf 87}, 174503 (2013).
\bibitem{GABB09} T. Goko, A.A. Aczel, E. Baggio-Saitovitch, S.L. Bud'ko, P.C. Canfield, J.P. Carlo, G.F. Chen, P. Dai, A.C. Hamann, W.Z. Hu, H. Kageyama, G.M. Luke, J.L. Luo, B. Nachumi, N. Ni, D. Reznik, D.R. Sanchez-Candela, A.T. Savici, K.J. Sikes, N.L. Wang, C.R. Wiebe, T.J. Williams, T. Yamamoto, W. Yu, and Y.J. Uemura, Phys. Rev. B {\bf 80}, 024508 (2009).
\bibitem{ABBC08} A.A. Aczel, E. Baggio-Saitovitch, S.L. Budko, P.C. Canfield, J.P. Carlo, G.F. Chen, P. Dai, T. Goko, W.Z. Hu, G.M. Luke, J.L. Luo, N. Ni, D.R. Sanchez-Candela, F.F. Tafti, N.L. Wang, T.J. Williams, W. Yu, and Y.J. Uemura, Phys. Rev. B {\bf 78}, 214503 (2008).
\bibitem{ZPYin11} Z. P. Yin, K. Haule, and G. Kotliar, Nat. Mater. {\bf 10}, 932 (2011).
\bibitem{ZPYin14} Z. P. Yin, K. Haule, G. Kotliar, Nat. Phys. {\bf 10}, 845 (2014).
\bibitem{Pratt} D. K. Pratt, W. Tian, A. Kreyssig, J. L. Zarestky, S. Nandi, N. Ni, S. L. Bud'ko, P. C. Canfield, A. I. Goldman, and R. J. McQueeney, Phys. Rev. Lett. {\bf 103}, 087001 (2009).
\bibitem{Christianson09} A. D. Christianson, M. D. Lumsden, S. E. Nagler, G. J. MacDougall, M. A. McGuire, A. S. Sefat, R. Jin, B. C. Sales, and D. Mandrus, Phys. Rev. Lett. {\bf 103}, 087002 (2009).
\bibitem{Hardy} F. Hardy, P. Adelmann, Th. Wolf, H. v. L$\rm \ddot{o}$hneysen, and C. Meingast, Phys. Rev. Lett. {\bf 102}, 187004 (2009).
\bibitem{KuFi14} H.-H. Kuo and I.R. Fisher, Phys. Rev. Lett. {\bf 112}, 227001 (2014).
\bibitem{WABC14} T.J. Williams, A.A. Aczel, S.L. Bud'ko, P.C. Canfield, J.P. Carlo, T. Goko, Y.J. Uemura, and G.M. Luke, arXiv:1408.3643 [cond-Mat] (2014).
\bibitem{WABB09} T.J. Williams, A.A. Aczel, E. Baggio-Saitovitch, S.L. Bud'ko, P.C. Canfield, J.P. Carlo, T. Goko, J. Munevar, N. Ni, Y.J. Uemura, W. Yu, and G.M. Luke, Phys. Rev. B {\bf 80}, 094501 (2009).
\bibitem{LGZL13} X. Lu, H. Gretarsson, R. Zhang, X. Liu, H. Luo, W. Tian, M. Laver, Z. Yamani, Y.-J. Kim, A.H. Nevidomskyy, Q. Si, and P. Dai, Phys. Rev. Lett. {\bf 110}, 257001 (2013).

\end{thebibliography}

\begin{thebibliography}{}

\bibitem{2bt7} J. W. Lynn, Y. Chen, S. Chang, Y. Zhao, S. Chi, W. Ratcliff, II, B. G. Ueland, and R. W. Erwin, Journal of Research of NIST {\bf 117}, 61-79 (2012).
\bibitem{2qhunag} Q. Huang, Y. Qiu, W. Bao, M. A. Green, J. W. Lynn, Y. C. Gasparovic, T. Wu, G. Wu, and X. H. Chen, Phys. Rev. Lett. {\bf 101}, 257003 (2008).
\bibitem{2xylu14} X. Y. Lu, J. T. Park, R. Zhang, H. Q. Luo, A. H. Nevidomskyy, Q. Si, and P. C. Dai, Science {\bf 345}, 657 (2014).
\bibitem{2man} H. R. Man, X. Y. Lu, J. S. Chen, R. Zhang, W. L. Zhang, H. Q. Luo, J. Kulda, A. Ivanov, T. Keller, E. Morosan, Q. Si, and P. C. Dai, Phys. Rev. B {\bf 92}, 134521 (2015).  
\bibitem{2XYLu15} X. Y. Lu, K. F. Tseng, T. Keller, W. L. Zhang, D. Hu, Y. Song, H. R. Man, J. T. Park, H. Q. Luo, S. L. Li, A. N. Nevidomskyy, and P. C. Dai, Phys. Rev. B {\bf 93}, 134519 (2016).
\bibitem{2MLCZ15} H. Man, X. Lu, J.S. Chen, R. Zhang, W. Zhang, H. Luo, J. Kulda, A. Ivanov, T. Keller, E. Morosan, Q. Si, and P. Dai, Phys. Rev. B {\bf 92}, 134521 (2015).
\bibitem{2TJFV13} M. Tomi{\'c}, H.O. Jeschke, R.M. Fernandes, and R. Valent{\'i}, Phys. Rev. B {\bf 87}, 174503 (2013).
\bibitem{2BoMe16} A. E. B$\rm \ddot{o}$mer and C. Meingast, Comptes Rendus Physique {\bf 17}, 90 (2016).
\bibitem{2FVBC10} R. M. Fernandes, L.H. VanBebber, S. Bhattacharya, P. Chandra, V. Keppens, D. Mandrus, M.A. McGuire, B.C. Sales, A.S. Sefat, and J. Schmalian, Phys. Rev. Lett. {\bf 105}, 157003 (2010).
\bibitem{2SoBK00} J. E. Sonier, J. H. Brewer, and R. F. Kiefl, Rev. Mod. Phys. {\bf 72}, 769 (2000).
\bibitem{WABC14} T. J. Williams, A. A. Aczel, S. L. Bud'ko, P. C. Canfield, J. P. Carlo, T. Goko, Y. J. Uemura, and G. M. Luke, arXiv:1408.3643 [cond-Mat] (2014).
\bibitem{2CCEP10} A. Cano, M. Civelli, I. Eremin, and I. Paul, Phys. Rev. B {\bf 82}, 020408 (2010).
\bibitem{2KACF12} H.-H. Kuo, J.G. Analytis, J.-H. Chu, R.M. Fernandes, J. Schmalian, and I.R. Fisher, Phys. Rev. B {\bf 86}, 134507 (2012).
\bibitem{2ACano12} A. Cano and I. Paul, Phys. Rev. B {\bf 85}, 155133 (2012).
\bibitem{2HuSK12} J. P. Hu, C. Setty, and S. Kivelson, Phys. Rev. B {\bf 85}, 100507 (2012).
\bibitem{2QDLR15} M. Qin, S. Dong, J. Liu, and Z. Ren, New J. Phys. {\bf 17}, 013011 (2015).
\bibitem{2BNNS09} S. L. Bud'ko, N. Ni, S. Nandi, G. M. Schmiedeshoff, and P. C. Canfield, Phys. Rev. B {\bf 79}, 054525 (2009).
\bibitem{2WaDN09} Y. Wang, Y. Ding, and J. Ni, Solid State Communications {\bf 149}, 2125 (2009).
\bibitem{2Paul14} I. Paul, Phys. Rev. B {\bf 90}, 115102 (2014).
\bibitem{2wien2k} K. Schwarz, P. Blaha, and G. K. H. Madsen, Comput. Phys. Commun. 147, 71 (2002).
\bibitem{2wien2wannier} Jan Kune, Ryotaro Arita, Philipp Wissgotte, Alessandro Toschie, Hiroaki Ikedaf, Karsten Helde, Computer Physics Communications {\bf 181}, Issue 11, 1888-1895 (2010).
\bibitem{2wannier90} A. A. Mostofi, J. R. Yates, Y.-S. Lee, I. Souza, D. Vanderbilt and N. Marzari, Comput. Phys. Commun. 178, 685 (2008).
\bibitem{2Sing08} D. J. Singh, Phys. Rev. B {\bf 78}, 094511 (2008).
\bibitem{2MaSc08} T. A. Maier and D. Scalapino, Physical Review B {\bf 78}, 020514 (2008).
\bibitem{2MGSH09} T. A. Maier, S. Graser, D. Scalapino, and P. Hirschfeld, Physical Review B {\bf 79}, 134520 (2009).
\bibitem{2MaHS12} T. A. Maier, P. J. Hirschfeld, and D. J. Scalapino, Physical Review B {\bf 86}, 094514 (2012).
\bibitem{2FeSc12} R. M. Fernandes, A. V. Chubukov, and J. Schmalian, Nat. Phys. {\bf 10}, 97 (2014).
\bibitem{2Miha11} B. Mihaila, arXiv:1111.5337 [cond-Mat, Physics:math-Ph] (2011).
\bibitem{2WKZB13} Y. Wang, A. Kreisel, V. Zabolotnyy, S. Borisenko, B. B$\rm \ddot{u}$hner, T. Maier, P. Hirschfeld, and D. Scalapino, Phys. Rev. B {\bf 88}, 174516 (2013).
\bibitem{2KOAU08} K. Kuroki, S. Onari, R. Arita, H. Usui, Y. Tanaka, H. Kontani, and H. Aoki, Phys. Rev. Lett. {\bf 101}, 087004 (2008).
\bibitem{2NoRo14} C. Noce and A. Romano, Phys. Status Solidi B {\bf 251}, 907 (2014).
\bibitem{2LTZL14} X. Lu, D.W. Tam, C. Zhang, H. Luo, M. Wang, R. Zhang, L. W. Harriger, T. Keller, B. Keimer, L.-P. Regnault, T. A. Maier, and P. Dai, Phys. Rev. B {\bf 90}, 024509 (2014).


\end{thebibliography}
\end{document}